\documentclass[conference]{IEEEtran}

\ifCLASSINFOpdf
\else
\fi

\usepackage[utf8]{inputenc}
\usepackage{graphicx}
\usepackage{dirtree}
\usepackage{color}   
\usepackage[export]{adjustbox}%
\usepackage{cite}
\usepackage{amsmath, amssymb}

\usepackage{amsthm}
\newtheorem{definition}{Definition}

\usepackage{xcolor}
\usepackage[colorlinks=false, linkcolor=black, citecolor=black, linktoc=all, hidelinks]{hyperref}

\usepackage[
    n,
    operators,
    advantage,
    sets,
    adversary,
    landau,
    probability,
    notions,
    logic,
    ff,
    mm,
    primitives,
    events,
    complexity,
    asymptotics,
    keys]{cryptocode}
\createprocedureblock{boxedprocedure}{center,boxed}{}{}{}
\newcommand{\revision}[1]{\textcolor{black}{{#1}}}

\usepackage[show]{chato-notes}
\usepackage{amsthm}
\usepackage{amssymb} 
\usepackage{pifont}
\usepackage{wrapfig}

\usepackage[frozencache,cachedir=.]{minted}

\usepackage[skip=2pt]{caption} %
\usepackage{subcaption}

\usepackage{cancel}
\usepackage{msc}
\usepackage{tikz}
\usetikzlibrary{matrix,shapes,arrows,positioning,chains, calc}

\usepackage[capitalize,noabbrev]{cleveref}
\usepackage{booktabs}
\usepackage[tikz]{mdframed}
\usepackage{xcolor}

\usepackage{hyperref}
\hypersetup{
    colorlinks=true,
}

\usepackage{tcolorbox}%
\definecolor{framecolor}{rgb}{0.122, 0.435, 0.698}%
\definecolor{bgcolor}{rgb}{0.95, 0.95, 0.95}%

\newcommand{\attackerbox}[1]{%
  \begin{tcolorbox}[
    colframe=framecolor,       %
    colback=bgcolor,           %
    boxrule=0pt,               %
    leftrule=3pt,              %
    arc=0pt,                   %
    left=2pt,                  %
    right=2pt,                 %
    top=2pt,                   %
    bottom=2pt,                %
    width=\linewidth,          %
    before skip=4pt,           %
    after skip=4pt             %
  ]
    #1
  \end{tcolorbox}
}

\newmdenv[
  linecolor=blue!60!black,
  backgroundcolor=blue!5!white,
  leftmargin=10pt,
  rightmargin=80pt,
  roundcorner=5pt,
  frametitle={Agent \A},
  frametitlebackgroundcolor=blue!15!white
]{personaframe}

\newmdenv[
  linecolor=teal!60!black,
  backgroundcolor=teal!5!white,
  leftmargin=80pt,
  rightmargin=10pt,
  roundcorner=10pt,
  frametitle={Agent \B},
  frametitlebackgroundcolor=teal!15!white
]{personbframe}

\usepackage{enumitem}
\usepackage{multirow}

\newcommand{\eg}{e.g.,\xspace}
\newcommand{\ie}{i.e.,\xspace}

\newcommand{\CR}[1]{\textcolor{black}{{#1}}}

\newcommand{\user}{\texttt{User}\xspace}
\newcommand{\agent}{\texttt{Agent}\xspace}
\newcommand{\provider}{\texttt{Provider}\xspace}
\newcommand{\ca}{\texttt{CA}\xspace}
\newcommand{\A}{\texttt{A}\xspace}
\newcommand{\B}{\texttt{B}\xspace}
\newcommand{\M}{\texttt{M}\xspace}
\newcommand{\U}{\texttt{U}\xspace}
\newcommand{\token}{\texttt{token}\xspace}

\newcommand{\uid}[1]{\texttt{uid}_{\texttt{#1}}}
\newcommand{\aid}[1]{\texttt{aid}_{\texttt{#1}}}

\newcommand{\name}[1]{\texttt{name}_{\texttt{#1}}}
\newcommand{\device}[1]{\texttt{device}_{\texttt{#1}}}
\newcommand{\IP}[1]{\texttt{IP}_{\texttt{#1}}}
\newcommand{\port}[1]{\texttt{port}_{\texttt{#1}}}

\newcommand{\ED}[1]{\texttt{ED}_{\texttt{#1}}}

\newcommand{\meta}[1]{\texttt{M}_{\texttt{#1}}}

\newcommand{\keygen}{\text{KeyGen}()\xspace}

\newcommand{\signing}[2]{\text{Sign}_{#1}\left(#2\right)}

\newcommand{\verification}[3]{\text{Verify}_{#1}\left(#2, #3\right)}

\newcommand{\signature}[2]{\sigma_{#1}^{#2}}

\newcommand{\PK}[1]{\texttt{PK}_{\texttt{#1}}}
\newcommand{\SK}[1]{\texttt{SK}_{\texttt{#1}}}

\newcommand{\PAC}[1]{\texttt{PAC}_{\texttt{#1}}}
\newcommand{\SAC}[1]{\texttt{SAC}_{\texttt{#1}}}

\newcommand{\OTK}[2]{\texttt{OTK}_{\texttt{#1}}^{\texttt{#2}}}
\newcommand{\SOTK}[2]{\texttt{SOTK}_{\texttt{#1}}^{\texttt{#2}}}

\newcommand{\crt}[1]{\texttt{Cert}_{\texttt{#1}}}

\newcommand{\myDH}[2]{\texttt{DH}_{\texttt{#1}}^{\texttt{#2}}}
\newcommand{\SDHK}{\texttt{SDHK}}

\newcommand{\nonce}{\texttt{N}}
\newcommand{\getnonce}{\nonce \xleftarrow{\$} \mathcal{R}}

\newcommand{\timestamp}[1]{\texttt{T}_{\texttt{#1}}}
\newcommand{\quota}[1]{\texttt{Q}_{\texttt{#1}}}

\newcommand{\myparagraph}[1]{\smallskip \noindent \textbf{#1.}}

\newcommand{\saga}{SAGA\xspace}

\newcommand{\messLabel}[1]{\small{#1}}
\definecolor{darkgray}{gray}{0.6}
\definecolor{gris}{gray}{0.85}

\definecolor{codebg}{rgb}{0.975,0.975,0.975 }

\newcommand{\acp}[1]{\texttt{CP}_{\texttt{#1}}}

\newcommand{\userregistry}{\mathcal{D}_U}
\newcommand{\agentregistry}{\mathcal{D}_A}

\newcommand\shortsection[1]{\vspace{3pt}{\noindent\bf #1.}}
\newcommand\shortersection[1]{\vspace{3pt}{\noindent\em #1.}}

\newcommand{\otkbudget}[2]{\text{Budget}_\texttt{OTK}(#1, #2)}

\newcommand{\otkcounter}[2]{\text{Counter}_\texttt{OTK}[#1][#2]}

\newcommand{\createcert}[2]{\text{GenCert}_{#1}(#2)}

\newif\ifdeitemized
\deitemizedfalse %

\newcommand{\paperlink}{\url{https://github.com/gsiros/saga}}

\begin{document}
\title{\saga: A Security Architecture for Governing\\ AI Agentic Systems}

\makeatletter
\newcommand{\linebreakand}{%
  \end{@IEEEauthorhalign}
  \hfill\mbox{}\par
  \mbox{}\hfill\begin{@IEEEauthorhalign}
}
\makeatother

\author{
    \IEEEauthorblockN{Georgios Syros}
    \IEEEauthorblockA{Northeastern University\\syros.g@northeastern.edu}
    \and
    \IEEEauthorblockN{Anshuman Suri}
    \IEEEauthorblockA{Northeastern University\\ans.suri@northeastern.edu}
    \and
    \IEEEauthorblockN{Jacob Ginesin}
    \IEEEauthorblockA{Northeastern University\\ginesin.j@northeastern.edu}
    \linebreakand
    \IEEEauthorblockN{Cristina Nita-Rotaru}
    \IEEEauthorblockA{Northeastern University\\c.nitarotaru@northeastern.edu}
    \and  
    \IEEEauthorblockN{Alina Oprea}
    \IEEEauthorblockA{Northeastern University\\a.oprea@northeastern.edu}
}

\maketitle

\begin{abstract}
Large Language Model (LLM)-based agents  increasingly  interact, collaborate, and delegate tasks to one another autonomously with minimal human interaction. 
Industry guidelines for agentic system governance emphasize the need for users to maintain comprehensive control over their agents, mitigating potential damage from malicious agents. Several proposed agentic system designs address agent identity, authorization, and delegation, but remain purely theoretical, without concrete implementation and evaluation. Most importantly, they do not provide user-controlled agent management.

To address this gap, we propose \saga, a scalable \textbf{S}ecurity \textbf{A}rchitecture for \textbf{G}overning \textbf{A}gentic systems, that offers user oversight over their agents' lifecycle. In our design, users register their agents with a central entity, the \provider, that maintains agent contact information, user-defined access control policies, and helps agents enforce these policies on inter-agent communication. We introduce a cryptographic mechanism for deriving access control tokens, that offers fine-grained control over an agent's interaction with other agents, providing formal security guarantees. 
We evaluate \saga on several agentic tasks, using agents in different geolocations, and multiple on-device and cloud LLMs, demonstrating minimal performance overhead with no impact on underlying task utility in a wide range of conditions. Our architecture enables secure and trustworthy deployment of autonomous agents, accelerating the responsible adoption of this technology in sensitive environments.\footnote{\CR{This is the full version of the paper accepted for publication at the Network and Distributed System Security Symposium (NDSS) 2026.}}

\end{abstract}

\IEEEpeerreviewmaketitle

\section{Introduction}
\label{sec:introduction}

AI agents with increased levels of autonomy are being deployed in safety-critical applications, such as healthcare \cite{mehandru2024evaluating, wang2025surveyllmbasedagentsmedicine}, finance \cite{wu2023bloomberggpt, nie2024surveylargelanguagemodels, zhou2024finrobotaiagentequity}, and cybersecurity \cite{zhang2025llms,rigaki2024hackphyrlocalfinetunedllm, kobayashi2025constructionevaluationllmbasedagents}. These agents, built on top of %
Large Language Models (LLMs), excel at automating complex tasks traditionally performed by humans. Agents powered by LLMs have sophisticated reasoning capabilities and the ability to understand and generate natural language. They also leverage access to tools installed on user devices, external resources, and the ability to interact with other AI agents autonomously. 

The increasing autonomy and functionality of AI agents expand the attack surface of agentic systems, introducing numerous security risks. As AI agents become more integrated into critical applications, securing these systems presents several challenges, as highlighted in a recent  OpenAI white paper \cite{shavit2023practices}.
Several key requirements include defining unique identities for AI agents, authenticating these agents, and providing secure mechanisms for agent discovery and communication.
These requirements must remain effective even under adversarial conditions, as malicious actors may attempt to impersonate agents, intercept communications, or manipulate agent behavior to extract sensitive information or subvert intended functionality.
Significantly, the OpenAI white paper emphasizes the necessity of maintaining user control and supervision throughout all operational phases and throughout the entire lifecycle of LLM agents to safeguard against potential harm from malicious agents.

Although various AI agentic system designs have been proposed that incorporate agent identities \cite{chan2024ids}, attribution \cite{chan2025infrastructure}, authorization mechanisms and delegation capabilities \cite{south2025authenticated}, these designs largely remain theoretical without implementation or evaluation. Most critically, they fail to adequately address the essential component of user-controlled agent management. Recently, Google's A2A protocol \cite{Surapaneni2025A2A} introduces a decentralized identity framework where agents advertise public metadata and initiate direct, encrypted communication using web-based authentication. While A2A promotes interoperability and supports verifiable identifiers, it lacks policy enforcement mechanisms and runtime mediation of agent interactions, and does not provide mitigation against adversarial agents.

\begin{figure*}[t]
    \centering
    \includegraphics[page=1,width=0.95\linewidth]{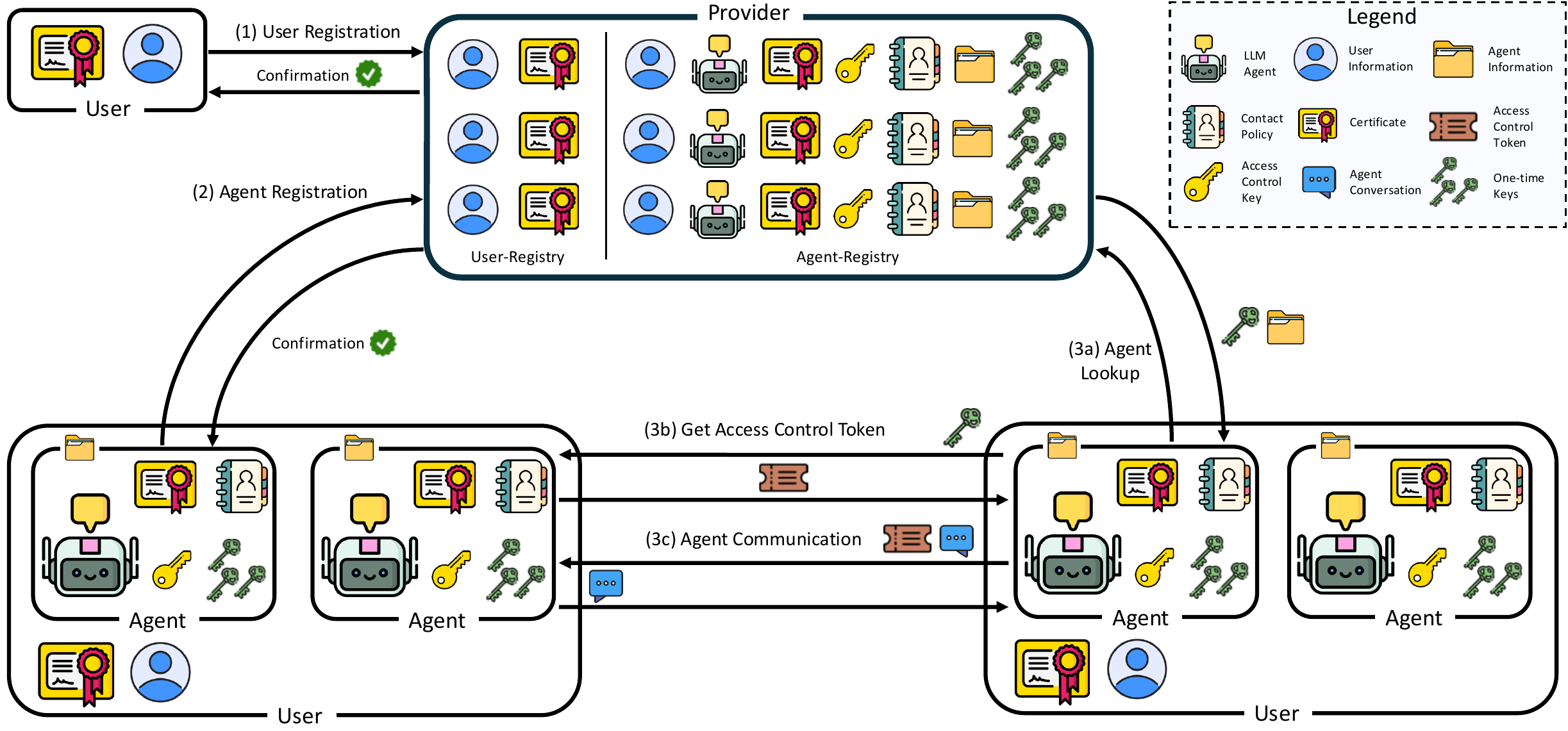} 
    \caption{System overview of \saga. (1) A user registers with the \provider. (2) A registered user registers their agent and receives a confirmation from the \provider. (3a) An initiating agent requests a One-time Key (OTK) and the receiving agent's information from the \provider. (3b) The initiating agent contacts the receiving agent with the OTK, after which the agents derive a shared key and the  receiving agent encrypts an access control token under the shared key. (3c) The initiating agent attaches this token to subsequent communication over TLS. When token expires or reaches limit, a new OTK is retrieved from the \provider\ and a new token is obtained from the receiving agent.}
    \label{fig:architecture_overview}
\end{figure*}

In this paper, we propose \saga, a framework for governing  LLM agent deployment, designed to enhance security while offering user oversight on their agents' lifecycle (see \Cref{fig:architecture_overview} for an overview). In \saga users register themselves and their agents with a \provider service that maintains user and agent metadata and  facilitates controlled communication establishment between agents. \saga enables users to control access to their agents through an Access Contact Policy that users  define for their agents. The enforcement of the policy is done through cryptographic access control tokens, derived from agent one-time keys  stored at the \provider.  Inter-agent communication over secure TLS channels does not involve the \provider for scalability, while utilizing access control tokens to ensure compliance with user-defined policies. The granularity of access control tokens provides a tradeoff between potential windows of vulnerability and performance overhead. 
The architecture aligns with best practices for securing agentic AI systems and meets key requirements outlined in emerging AI governance frameworks.

\myparagraph{Contributions} We highlight our main contributions:
\begin{itemize}
\item We present \saga, a \textbf{S}ecurity \textbf{A}rchitecture for \textbf{G}overning \textbf{A}gentic systems, that enables users control and oversight of their LLM-enabled agents. 
\item We propose a cryptographic mechanism based on access control tokens encrypted under shared agent keys that allows fine-grained control over inter-agent communication to balance security and performance considerations. 

\item \revision{We use PROVERIF to formalize the secrecy of the SAGA token, authentication of communication between agents and the \provider, and authentication of communication between any two agents. We automatically prove these properties with respect to an attacker that can observe, intercept, modify, replay, reorder, and synthesize arbitrary messages on the network.}

\item We evaluate our system across several agentic tasks (scheduling meetings, submitting expense reports, and collaborative writing), multiple on-device and cloud LLM models, and agents placed in diverse geolocations to demonstrate the scalability of our design.

\item \revision{We discuss and evaluate the fault-tolerance and scalability of the \provider. Our results demonstrate that making the \provider fault-tolerant using RAFT introduces negligible throughput degradation across key operations. They also demonstrate that scaling the \provider using sharding  results in linear throughput increase.}
\end{itemize}
\CR{The code of our protocol and the formal verification is available at \paperlink.}

\section{Background and Problem Statement}
We provide background on agentic systems, their security requirements, and define our problem statement.

\subsection{Agentic AI Systems}
Agentic AI systems represent an advancement in autonomous systems, enabled by generative AI. These agents independently devise execution plans for  tasks, leveraging their underlying models for strategic planning and reasoning. As part of the planned steps for completing the assigned tasks, agents might need to leverage additional tools installed on user devices, interact with cloud services, or communicate with external agents running on other devices.
LLM agents can automate both professional tasks like scheduling and personal activities such as trip planning. For example, Alice's calendar agent might contact Bob's calendar agent to arrange a meeting, then use an email tool to send confirmations. We formulate the following definitions and terminology that we use in the rest of the paper.
\begin{definition}
A \user  owns one or more agents and is responsible for assigning them tasks.  
A user has only one  agent instance running on each device (\eg only one Calendar agent running on a particular device).
\end{definition}
\begin{definition}
An \agent is an autonomous software entity using  LLMs for decision making. Upon receiving a user query, agents use LLMs for planning, store information in memory, and execute plans to complete assigned tasks. Agents can access tools installed on user machines. Each agent operates under a specific user's authority and may interact with other agents to fulfill complex objectives. \revision{Agents can be "initiating" (starting communication) or "receiving" (accepting communication) with other agents.}
\end{definition}
\begin{definition}
A task is defined as a sequence of actions that an \agent must plan and execute to achieve a \user-defined goal. Actions might involve interacting with an LLM, invoking other tools, or communicating with other agents.
\end{definition}
\subsection{Requirements \revision{and Challenges} for Secure Agentic Systems}

An OpenAI white paper \cite{shavit2023practices} outlines several guidelines and open problems in designing secure agentic systems. Key requirements include defining unique identities for AI agents with verification capabilities, providing secure mechanisms for agents to discover and communicate with each other, and enabling agents to make informed decisions about resource access and collaboration with other agents.  
Agentic systems must also implement protections against adversarial or rogue agents to limit their influence and prevent harm to benign agents. 
Crucially, \cite{shavit2023practices} advocates for users' control and oversight throughout the lifecycle of LLM agents.

\revision{\textbf{Challenges}:} \revision{\CR{An agent management system must address three key challenges: (1) \textit{discovery}---how agents discover each other; (2) \textit{secure communication}---how agents communicate with each other; and (3) \textit{remote access control}---which agents are allowed to interact with a specific agent, for what tasks, and for how long.} While inter-agent communication should be performed automatically by agents, access control and agent discovery should remain \CR{under the user's control} for better security. Existing solutions address only one of these challenges and are not specifically designed for agents. For example, secure messaging protocols like Signal or Matrix provide peer discovery and secure communication, but lack fine-grained access control; protocols like TLS or QUIC enable  secure communication, but offer no agent discovery or fine-grained access control; and systems like Kerberos provide high-granularity access control, but lack secure communication capabilities. The recent A2A protocol~\cite{Surapaneni2025A2A} is the only agent-specific work we are aware of that ensures communication between agents, but provides no security guarantees or access control functionality.}

\subsection{Problem Statement}

\revision{Our goal is to design a system that addresses all the above challenges by providing agent discovery, secure agent communication, and fine-grained access control to the services provided by an agent,  while allowing users to retain control over the management of their agents.}

We  first address the fundamental requirement of creating and managing agent identities while providing effective mechanisms for agent discovery within the ecosystem. We  design secure inter-agent communication protocols, which represents a critical functionality of any agentic system, and address the challenge of how to limit the influence of  adversarial agents on benign ones. We also aim to enable user control on an agent's entire lifecycle, including agent registration in the system, managing agent's contacts, and  agent's  deactivation at any time, according to the user's preferences and policies. 

With the rapid evolution of agentic systems and emerging regulations, our goal is to design a modular, extensible framework that can support new security properties with minimal changes to core components.

\section{System Architecture}

We begin by outlining the desirable goals for a secure agentic system (\Cref{sec:system_goals}), followed by an overview of our proposed architecture \saga (\Cref{sec:arch_overview}). We also discuss the system assumptions  (\Cref{sec:system_assumptions}) and the considered threat model (\Cref{sec:threat_model}).

\subsection{System Goals}
\label{sec:system_goals}

Although the requirements for secure agentic systems are still evolving, we outline several desirable goals:

\shortsection{Agent lifecycle managed by users} %
\CR{An agent’s lifecycle should be established by a user. The user installs the agent on their device, registers it in the system, and retains the authority to terminate the agent at any time. Termination is performed by deactivating the agent and preventing other agents from locating it in the ecosystem.}

\shortsection{Agent access controlled by user policies} While LLM agents can autonomously decide to contact other agents to perform user-assigned tasks, communication between agents should be controlled by user-defined access control policies. 

\shortsection{Limited trust}
Agents should have limited trust in other third-parties in the system, including any centralized service provider or other agents. 

\shortsection{Scalability}
Design the system to efficiently manage a large number of agents  with minimal performance overhead.

\shortsection{Limited influence on other agents} An agent can control only its own participation in the system and cannot arbitrarily remove other agents from the ecosystem.

\shortsection{Limited vulnerability window}
It is crucial to limit the vulnerability window when compromised agents are still active in the system. The length of this  window can be controlled by cryptographic mechanisms and depends on  factors such as the agent’s communication patterns, the nature of the task, and the sensitivity of the exchanged data. 

\shortsection{Maintain system utility}
We aim to preserve the utility of the system, as measured by task completion, without compromising system security. Additionally, the framework should be agnostic to various agent implementations and LLM models that serve as the backbone.

\subsection{Architecture Overview}
\label{sec:arch_overview}

We provide an overview of our \saga system architecture in \Cref{fig:architecture_overview}.
A critical consideration in designing a secure agentic framework is the mechanism for agent discovery. The implementation of an agent registry facilitates this discovery process, with architectural options ranging from centralized to distributed registry models. In our design, we choose to use a centralized registry, maintained by an entity called \provider.
The \provider manages agent and user identities, ensures authentication, and facilitates controlled communication establishment between agents.

To ensure user  oversight over an agent's lifecycle, we delegate the agent registration process to users themselves, with the \provider maintaining the agent  information. Users must be registered in the system and authenticate before registering any agents and retain the ability to  deactivate their agents, for example as a protective measure when abuse is detected. Each agent is assigned a unique identifier, linked to the user and the device it is installed on. The \provider thus maintains the \texttt{User Registry} ($\userregistry$)  holding user identity records, and the \texttt{Agent Registry} ($\agentregistry$)  storing agent metadata, cryptographic credentials, endpoint information, and agent communication policies.

While communication between agents can be implemented using standard secure protocols such as TLS, we would like the ability to enforce limits on agent access, controlled by users.
Governed by the same principle of users managing  access to their agents, each user can specify an \emph{Access Contact Policy} for each agent defining which  agents are permitted to initiate contact.
To enable fine-grained access control and limit the vulnerability window, access to agents needs to be cryptographically bounded. 
For scalability and reduced reliance on the \provider, once an agent obtains the necessary connection details for another agent, all subsequent communication should proceed directly, without \provider involvement.

An important design consideration is how to manage agent access control in a fine-grained manner while reducing the burden on the \provider. Towards this goal, each  agent registers a number of public one-time keys (OTKs) with different quotas for each initiating agent (and locally stores the corresponding secret information linked to these $\OTK{}{}$s).
In principle, an $\OTK{}{}$ could be used only once for an inter-agent interaction, but that would require generating and storing a large number of $\OTK{}{}$s, presenting significant operational overhead. We extend the lifetime of these one-time keys by the receiving agent 
generating an \emph{Access Control Token} that is encrypted using a dynamically derived shared key between the agents.
The shared key must be unique to each initiating–receiving agent pair and derivable only by those two agents.
To support the shared key computation, each agent maintains a long-term \emph{Access Control Key}, which serves as the basis for deriving the shared key and binding it to the specific agent pair.
The public component of the access control key is stored in the Agent Registry, while each agent maintains the secret part.
We then use the Diffie-Hellman key exchange protocol, in which the initiating agent contributes the access control key, and the receiving agent the $\OTK{}{}$s as public keys. The receiving agent encrypts the \emph{Access Control Token} under the derived shared key.
The token includes an expiration timestamp and a limit on the number of permitted requests.
The token is reused for inter-agent communication, without involving the \provider.
When the token expires, a new one is created after the initiating agent obtains another $\OTK{}{}$ from the \provider. This design balances the security and performance overhead in the system, by allowing users to tune access to their agents through the number of keys they register and the lifetime of \emph{Access Control Tokens}. Note that the protocol is asymmetric, as access control is enforced for receiving agents, according to user-specified policies for initiating agents.

Below, we outline the main components of our system; protocol details are provided in \Cref{sec:protocol}.

\shortsection{User Registration (\Cref{sec:user_registration})}
Users must register with the \provider using a persistent identity mechanism, such as OpenID Connect \cite{openidconnect}. Upon successful registration, users authenticate to the \provider and provision agents for participation in the \saga ecosystem.

\shortsection{Agent Registration (\Cref{sec:agent_registration_by_user})}
A user registers its agents with the \provider. During registration, the user generates cryptographic keys for its agents, including TLS certificates and access control keys. The user also signs the agent metadata, such as hostname, port, and device identifiers, to bind it to its identity and specific device. Additionally, the \provider signs the agent’s metadata.

\shortsection{Agent Management (\Cref{sec:agent_management})} 
The \provider enables users to define and update an \emph{Agent Contact Policy} for each of their agents, which governs the policy for permissible incoming communication.
This policy allows users to restrict which other agents can initiate contact with their agent, and to impose access control constraints.
The \provider enforces these policies during initial contact requests, ensuring that agent interactions are governed by user-defined rules. Users can deactivate their own agents at any time, but cannot deactivate agents owned by others.

\shortsection{Agent Communication (\Cref{sec:agent_communication})} To initiate contact with another agent, the initiating agent queries the \provider with the receiving agent's identifier. The \provider responds with metadata, including the receiving agent's device, IP address, and a one-time key ($\OTK{}{}$) for access control. Subsequently, the initiating agent establishes a shared key with the receiving agent using the Diffie-Hellman protocol. 
The receiving agent generates an access control token, encrypted under the shared key, which is included in any communication by the initiating agent. When a token expires or reaches its request limit, the initiating agent obtains a new $\OTK{}{}$ from the \provider. Note that the \provider does not mediate inter-agent communication, but it is critical in enforcing each receiving agent's Access Control Policy by distributing $\OTK{}{}$s to initiating agents.

\subsection{System Assumptions}
\label{sec:system_assumptions}

The correctness and security guarantees of \saga rely on a set of clearly defined system assumptions, outlned below.

\shortsection{Secure User Authentication and Human Verification} We assume that the \provider implements a robust user authentication mechanism (\eg OpenID Connect) and that user credentials are not compromised. Crucially, we assume that agent registration is restricted to authenticated human users, enforced through human verification during user account creation. This verification process is delegated to a trusted external identity service, which certifies the user’s human status on behalf of the \provider. 

\shortsection{Agent Identity Control} We assume that attackers can create and register agents under their own identities but cannot impersonate other users. That is, while adversaries may instantiate and operate malicious agents, they cannot register agents under the identity of a benign user.

\shortsection{Public IP Addressing} All agents and providers are assumed to operate under globally routable, public IP addresses. This design avoids NAT traversal and local discovery, assuming that agents are reachable at their registered endpoints.

\shortsection{Cryptographic Soundness} We assume that all cryptographic primitives used by the system---signature schemes, key exchange protocols, encryption schemes, and key derivation functions---are secure. 
Secret keys are assumed to remain confidential and outside adversarial control.

\shortsection{Secure Channels} All communication, both between agents and between agents and the \provider, is protected by TLS.
We assume that TLS provides confidentiality, integrity, and authenticity against network-level adversaries performing  message tampering, eavesdropping, and replay.

\shortsection{Network Protections} We assume that the network infrastructure enforces basic protections against denial-of-service attacks and packet flooding. %

\subsection{Threat Model}
\label{sec:threat_model}

The \provider is expected to adhere to the \saga protocol logic, including enforcing contact policies, issuing keys, and performing registry operations. However, it may be \emph{honest-but-curious}:  capable of observing agent metadata and traffic patterns without actively attempting to subvert the protocol. The user and agent registries are assumed to be securely stored and not vulnerable to adversarial control or tampering.
We consider several adversarial capabilities: 

\noindent \textbf{C1}: Adversaries might create agents and register them with the \provider. These adversarial agents could deviate from the protocol when communicating with other agents. They could also add themselves to the contact policy of benign agents by performing social engineering on users.

\noindent \textbf{C2}: A legitimate agent registered with the \provider could be compromised by an adversary. This attack could occur when agents interact with external resources, such as websites, or tools installed on user devices, which might trigger a compromise.   

\noindent \textbf{C3}: Adversaries may instruct an agent to self-replicate on the same device or on another user's device without registering the child agent with the \provider. Prior work has demonstrated such self-replication of agents \cite{cohen2024here}. The parent agent can share TLS keys, access control keys, and existing access control tokens with the child agent.

\noindent \textbf{C4}: An adversarial agent may share its TLS public keys, access control keys, and access control tokens with another adversary-controlled agent, enabling communication with a benign victim agent.

\noindent \textbf{C5}: An adversary could attempt to mount a Sybil attack, by creating agents with multiple identities. 

\noindent \revision{\textbf{C6}:} \revision{An adversary may overhear, intercept, and synthesize any message, and is only limited by the computational hardness constraints of the employed cryptographic primitives.}

\section{\saga\ Protocol Specification}
\label{sec:protocol}

In this section, we begin with a description of the cryptographic primitives involved in our protocol, followed by a description of the key protocols involved in \saga: user registration (\Cref{sec:user_registration}), agent registration (\Cref{sec:agent_registration_by_user}), agent management (\Cref{sec:agent_management}), and inter-agent communication (\Cref{sec:agent_communication}).

\subsection{Cryptographic Primitives and Notation}
We leverage the following cryptographic primitives:

\shortersection{Signature schemes} A signature scheme consists of three algorithms: $\keygen$ -- a key generation function that outputs a (public, private) signing key pair \revision{$(\PK{},\SK{})$},
   
$\signing{\SK{}}{m}$ -- a signing algorithm that outputs a signature $\sigma$ on message $m$ using $\SK{}$, and $\verification{\PK{}}{m}{\sigma}$ -- an algorithm that  verifies if the signature $\sigma$ on message $m$ is correct.
We assume that the signature scheme is Existential Unforgeable under Chosen Message Attack \cite{Katz_Lindell}, such as ECDSA \cite{johnson2001elliptic} and Ed25519 \cite{rfc8709}.
A certificate generation function $\createcert{X}{m}$ involves entity $X$ creating a certificate for content $m$ as: $\createcert{X}{m} = \langle m, \signing{\revision{\SK{X}}}{m} \rangle$

\shortersection{Hash function} We use a collision-resistant hash function $\texttt{H}(\cdot)$, such as SHA-256 or SHA-3 \cite{dworkin2015sha}. 

\shortersection{Diffie-Hellman Key Exchange} The Diffie-Hellman Key Exchange protocol \cite{Katz_Lindell} is a cryptographic method that allows two parties to establish a shared secret key. Each party generates a (secret, public) key pair, and exchanges with the other party the public component. 
We denote by $DH$ the function that takes as input the secret key of one party and public key of the other party and computes the shared secret key $DH(x,g^y)=DH(y,g^x) = g^{xy} \mod p$. The security of Diffie-Hellman is based on the Computational Diffie-Hellman (CDH) assumption. 

\shortersection{Key Derivation Function} A Key Derivation Function (KDF) is a cryptographic algorithm that derives one or more secret keys from a master secret. We use the HMAC-based Extract-and-Expand Key Derivation Function (HKDF) \cite{rfc5869} with SHA-256 as the underlying hash function.

\shortersection{Notation}
We introduce formal notation  throughout the protocol specification. For convenience, \Cref{tab:symbols} provides a summary of all symbols used.

\begin{table}[h]
    \centering
    \begin{tabular}{l|l}
    \toprule
    \textbf{Symbol} & \textbf{Description} \\
    \midrule
    $\PK{}, \SK{}$ & Public/private signing keys \\
    $\crt{}$ & Certificate issued by CA \\
    \midrule
    $\uid{U}$ & User identifier \\
    $\aid{A}$ & Agent \A identifier \\
    $\ED{A}$ & Agent \A endpoint descriptor \\
    $(\PK{A}, \SK{A})$ & Agent \A public/private TLS credentials \\
    $(\PAC{A}, \SAC{A})$ & Agent \A public/private access control keys \\
    $(\OTK{A}{i}, \SOTK{A}{i})$ & Agent \A one-time public/private keys \\
    $\acp{A}$ & Contact policy of agent \A \\
    \midrule
    \token & Access control token \\
    $\signature{X}{Y}$ & Entity Y-issued signature regarding entity X. \\
    \midrule
    $\userregistry$ & Provider’s user registry \\
    $\agentregistry$ & Provider’s agent registry \\
    \bottomrule
    \end{tabular}
    \caption{Notation used in the \saga protocol.
    }
    \label{tab:symbols}
\end{table}

\subsection{User Registration}
\label{sec:user_registration}

The first step for any user seeking to deploy agents is to register with the \provider. This process establishes the user's identity and enables them to link future agents to their identity and manage them securely. The user obtains a certificate signed by an external certificate authority \ca on their public key, sent to the \provider. 
We assume that the user can establish a TLS connection with the \provider and that the provider can verify the user's identity with the help of an external service such as OpenIDConnect. The protocol follows these steps (also shown in \Cref{fig:user_registration}):

\begin{enumerate}[leftmargin=*]

    \item \textbf{User Account Setup.}
    The user selects a public identifier $\uid{U}$ corresponding to their email address, \eg $\uid{U} = \texttt{`alice@domain.com'}$, and a secret passphrase \texttt{passwd} to authenticate to the \provider.

    \item \textbf{User signature key generation:}  
    The user generates a signature key pair
    $
    (\PK{U}, \SK{U})$ 
    for signing agent information. %
    
    The user contacts  \ca to generate its certificate $\crt{U} = \createcert{\SK{CA}}{\langle \uid{U}, \PK{U} \rangle}$, and shares it with the \provider.

    \item \textbf{Connection establishment.} The user obtains the \provider's certificate and public key $\PK{Prov}$ from the \ca and verifies them.
    A TLS session is established between the user and the \provider.

    \item \textbf{Sending user information.}
    The user submits $(\uid{U}, \texttt{passwd})$ and $\crt{U}$ to the \provider.

    \item \textbf{User identity verification.} The \provider verifies the user's identity using an external service \emph{S} (see \Cref{sec:system_assumptions}). If  verification is successful and the account does not exist, the \provider finalizes the  registration.
   
    \item \textbf{User account storage and confirmation:}  
    The \provider updates the user registry:  $\userregistry[\uid{U}] = \langle \texttt{H(passwd)}, \crt{U} \rangle$ and sends a confirmation to the user.

\end{enumerate}
After the user registration is completed successfully, the user can proceed to register its agents with the \provider.

\subsection{Agent Registration by User}
\label{sec:agent_registration_by_user}

The agent registration process ensures that each agent is cryptographically bound to its user and a specific user's device. 
The cryptographic information generated by a user for its agents is stored by the \provider and subsequently used to establish secure agent communication mediated by the \provider.
The following protocol is executed by user \U to register their agent \A (shown in \Cref{fig:agent_registration}, Appendix).

\begin{enumerate}[leftmargin=*]
 
    \item \textbf{Generating agent information}. The user selects an identifier $\name{A}$ for the agent, forming a unique agent ID in combination with their username:
    $
        \aid{A} = \uid{U}\texttt{:}\name{A}.  
    $
    The user specifies the agent's device name $\device{A}$ and networking details $\IP{A}$ and $\port{A}$.
    These comprise the agent's ``endpoint descriptor'':
    \begin{align*}
        \ED{A} = \langle \device{A}, \IP{A}, \port{A} \rangle
    \end{align*}

    \item \textbf{Generating cryptographic keys}. The user generates the following keys for the agent:\label{step:agent_registration_genkey}
    \begin{itemize}
        \item TLS credentials $(\PK{A}, \SK{A})$ to establish secure communication channels with other agents and a signed certificate by the \ca:
        \begin{align*}
            \crt{A} & = \createcert{\SK{CA}}{\langle \aid{A}, \PK{A} \rangle}
        \end{align*}
        \item A public-private key pair: $(\PAC{A}, \SAC{A})$ for access control.
        These  \CR{long-term} keys  are used for the receiving agent to derive a shared key with the initiating agent for encrypting the access control token in inter-agent communication (\Cref{sec:agent_communication}).%
                
        \item A batch of $N$ one-time public and secret key pairs $(\OTK{A}{1}, \SOTK{A}{1}), \dots, (\OTK{A}{N}, \SOTK{A}{N})$, used for deriving tokens for controlling access to the agent.
        Each public one-time key
        is signed by the user:
        \begin{align*}
            \signature{\OTK{}{i}}{\U} & = \signing{\SK{U}}{\langle \aid{A},\OTK{A}{i} \rangle}
        \end{align*}
    \end{itemize}

    The user also signs the agent's device and networking information, along with its long-term  keys:
    $$
        \signature{\A}{\U} = \signing{\SK{U}}{
            \langle \aid{A}, \ED{A}, \PK{A}, \PAC{A}, \PK{Prov} \rangle
        }
    $$

    The inclusion of $\PK{Prov}$ in the signature indicates that the agent is registered with the specified \provider.

\item \textbf{Specifying the contact policy}: 
 The user  specifies the agent's contact policy $\acp{A}$ (discussed in \Cref{sec:agent_management}).

    \item \textbf{User authentication to \provider}. The user establishes a secure TLS connection with the \provider, and
    authenticates by providing credentials $\langle \uid{U}, \texttt{passwd} \rangle$. The \provider verifies the credentials and proceeds if successful.

    \item \textbf{Registration  submission}. The user submits to the \provider: the agent’s information $(\aid{A}, \ED{A}, \acp{A})$, the TLS certificate  $\crt{A}$, the public access control and one-time keys $\PAC{A}, \{\OTK{A}{1}, \texttt{...}, \OTK{A}{N}\}$, and signatures $\signature{\A}{\U}, \quad \signature{\OTK{}{i}}{\U} \quad \text{for } i \in [1, N]$.
    
    The agent stores locally all the private keys corresponding to the public keys  submitted to the provider: ($\SK{A}$, $\SAC{A}$, $\{\SOTK{A}{1}, \texttt{...}, \SOTK{A}{N}\}$).

    \item \textbf{Provider verification}. The \provider processes the registration request by checking that $\aid{A}$ and $\ED{A}$ are globally unique,  and verifying  $\crt{A}$ and signatures:
    $$\text{Verify}_{\PK{U}}( 
            \langle \aid{A}, \ED{A}, \PK{A}, \PAC{A}, \PK{Prov} \rangle, \signature{\A}{\U})$$
        $$\text{Verify}_{\PK{U}}(\langle \aid{A}, \OTK{A}{i} \rangle, \signature{\OTK{}{i}}{\U})$$

    \item \textbf{Completion.} 
    Upon successful verification, the \provider stores in the agent registry the agent’s metadata $\meta{A}$, the contact policy $\acp{A}$, the agent's signatures $\signature{\A}{\U}$ and $\signature{\OTK{}{i}}{\U}$, along with $\uid{U}$ to associate the agent \A with user \texttt{U}:
    \begin{align*}
        & \meta{A} = \{\ED{A}, \crt{A}, \PAC{A},  \OTK{A}{i}, i \in [1,N] \} \\
        & \agentregistry[\aid{A}] = \langle \uid{U}, \meta{A}, \acp{A}, \signature{\A}{\U}, \signature{\OTK{}{i}}{\U} \rangle
    \end{align*}
    The \provider then signs the agent \A's information
    \begin{align}
        \signature{\A}{Prov} = \signing{\SK{Prov}}{\langle \aid{A}, \crt{A}, \ED{A}, \PAC{A}, \signature{A}{\U} \rangle} \nonumber
    \end{align}
    and returns it as confirmation to the user. The user stores this signature, which it uses when initiating agent communication (\Cref{sec:agent_communication}).    
    \label{step:agent_registration_ack}
\end{enumerate}
The agent \revision{\A} is now officially registered and can securely communicate within the ecosystem.

\subsection{Agent Management}
\label{sec:agent_management}

Agent management in \saga involves managing access control polices, policy updates, and cryptographic key management. 

\shortsection{Access control management}
There are two fundamental capabilities in \saga for managing access control: (1) specification and enforcement of the \emph{Agent Contact Policy} ($\acp{}$), and (2) provisioning of fresh \emph{One-Time Keys} ($\OTK{}{}$) to facilitate user-controlled  communication.

\subsubsection{Agent Contact Policy} 
\label{sec:agent_contact_policy}

In \saga, each agent is associated with a contact policy $\acp{}$ that specifies which initiating agents are authorized to establish contact. This policy is defined by the user when registering their agent, and is enforced by the \provider during contact resolution.
$\acp{}$ consists of a set of declarative rules, along with the number of $\OTK{}{}$s that should be allotted to an agent that matches that rule. These rules may use pattern matching over agent identifiers to allow flexible yet controlled specification of acceptable contacts.
For instance, the rule \texttt{("*@company.com:email\_agent", 10)} permits any email-handling agent from a specified domain to initiate contact and be allotted at most 10 $\OTK{}{}$s.

If multiple rules match, the one with the \emph{highest specificity} is selected. 
For instance, in \Cref{list:agent_contact_policy}, agent \texttt{alice@company.com:calendar\_agent} attempts to contact another agent. The agent's identifier matches the top three  patterns, but the first rule is the most specific and therefore determines the $\OTK{}{}$ budget, resulting in 15 $\OTK{}{}$s.

\begin{listing}[h]
\begin{minted}[fontsize=\scriptsize, bgcolor=codebg]{json}
// Agent C's Contact Policy
[
  {
    "agents": "alice@company.com:calendar_agent",
    "budget": 15
  },
  {
    "agents": "*@company.com:calendar_agent",
    "budget": 10
  },
  {
    "agents": "bob@mail.com:*",
    "budget": 100
  }
]
\end{minted}
\caption{Example contact policy for an agent. More specific patterns take precedence over general ones.}
\label{list:agent_contact_policy}
\end{listing}

Formally, for a receiving agent \( \A \) with contact policy \( \acp{A} \) and an initiating agent \( \B \) with identifier \( \aid{B} \), the number of $\OTK{}{}$s issued is defined as:
\begin{align}
    \otkbudget{\aid{A}}{\aid{B}} =
    \begin{cases}
        -1 & \text{if } \mathcal{R} = \emptyset \\
        \texttt{B}(r^*) & \text{if } \mathcal{R} \neq \emptyset \\
    \end{cases} \nonumber,
\end{align}
where $r^*$ is the most specific rule among all rules $\mathcal{R}$ in $\acp{A}$ that match $\aid{B}$, and $\texttt{B}(r^*)$ corresponds to the budget for rule $r^*$ indicated in $\acp{\A}$. The distinction for $\mathcal{R} = \emptyset$ helps the user \CR{distinguish between no match in policy and an expired $\OTK{}{}$ budget}.

\subsubsection{One-Time Key Generation}
To control communication to registered agents, the \provider facilitates the distribution of $\OTK{}{}$s: one-time keys generated and uploaded by each user for their agents. Each $\OTK{}{}$ is consumed when an initiating agent contacts the \provider to obtain information for a receiving agent. $\OTK{}{}$s are defined for receiving agents, as the protocol is asymmetric.
Additionally, the \provider maintains the number of $\OTK{}{}$s  remaining for a particular agent communicating with another agent. 
\CR{While it might be possible for the initiating agent to receive multiple $\OTK{}{}$s from the \provider, thereby reducing the number of times the agent must contact the \provider, we chose a more conservative design. In our approach, the agent receives only one $\OTK{}{}$ per request to limit the vulnerability window in case the agent is compromised.}

When an initiating agent \B queries the \provider to contact a recipient agent \A, the \provider first verifies that the initiating agent satisfies \A's Agent Contact Policy ($\acp{A}$), as detailed above.
If this is the first time \B is contacting \A, the \provider creates a counter $\otkcounter{\aid{A}}{\aid{B}}$ to keep track of the number of remaining $\OTK{}{}$s, and initializes it with $\otkbudget{\aid{A}}{\aid{B}}$.

If the policy check succeeds and a valid $\OTK{}{}$ is available (indicated by a positive value for $\otkcounter{\aid{A}}{\aid{B}}$), the \provider returns an $\OTK{}{}$ to the initiating agent (along with the recipient's metadata and its signature, discussed in \Cref{sec:agent_communication}) and decreases the counter by one.

Obtaining an $\OTK{}{}$ at this stage may fail due to: (a) exhaustion of $\aid{A}$'s overall $\OTK{}{}$ pool, or (b) depletion of $\aid{B}$'s $\OTK{}{}$ quota as defined by $\acp{A}$. The user of agent \A can update the contact policy and refresh the $\OTK{}{}$s at any time.

\shortsection{Policy Updates and Revocation}
Users can dynamically update their agents' contact policies via the \provider interface. Updates may include adding rules (\eg to onboard collaborators) or removing them (\eg to revoke access), allowing policies to adapt as trust relationships shift or threats emerge.

It is important to allow users to block contact from specific agents, as they might detect abusive behavior. To block specific agents, a receiving agent can update its contact policy with a rule that assigns a $\texttt{B}(\cdot)$ score of~$-1$.
The modified $\acp{}$ is then pushed to the \provider. %
Alternatively, to completely disable incoming contact, a user can request the \provider to deactivate their agent at any time. \CR{This functionality is motivated by our design goal of giving the user full control over their agent’s entire lifecycle.} Crucially, we do not allow users the ability to deactivate agents registered by other users, as they should control only their own agents.

\shortsection{Cryptographic key management} Cryptographic keys for agents should be managed by users according to best practice principles for key management and key rotation \cite{NIST80057r5}. In particular, users should periodically rotate their agents' TLS key and access control keys. Best practices should also be followed by the \provider for user authentication and password management \revision{\cite{grassi2017digital}}.

\subsection{Inter-Agent Communication}
\label{sec:agent_communication}
We describe how two registered agents can communicate with each other securely, while respecting the Access Contact Policy defined by users for their agents.

When an agent \B (initiating agent) wants to contact agent \A (receiving agent), it first queries the \provider to verify \A's registration. If permitted by \A's contact policy, \B receives \A's metadata and a one-time key ($\OTK{}{}$) from the \provider.
This $\OTK{}{}$ is used by both agents to derive a shared key, which the receiving agent \A will use to encrypt an access control token (ACT) for \B.
The ACT is scoped to a specific task, but the granularity of tokens can be adjusted further if desired. The token will have limited validity, as well as a limit on the total number of requests.  

The following steps outline the agent communication protocol involving initiating agent \B contacting receiving agent \A to obtain an access control token (shown in \Cref{fig:agent_communication}).
This protocol runs either the first time  \B contacts \A, or when  tokens have expired or exceeded their usage limits.

\begin{enumerate}[leftmargin=*]
    \item \textbf{Establishing a TLS connection with the \provider:}  
    This step follows a standard TLS session establishment between \B and the \provider.
    \label{step:agent_comm_tls_establish}
    \item \textbf{Receiving agent information retrieval:}  
    \B requests permission to contact \A by specifying their identity ($\aid{B}$) and the identity of the receiving agent ($\aid{A}$). The \provider verifies that \B is in \A's contact policy and has sufficient $\OTK{}{}$s allotted to it (by making sure $\otkcounter{\aid{A}}{\aid{B}} > 0$). It returns \A's access information: user's certificate $\crt{U1}$, agent's device and network information $(\aid{A}, \ED{A})$, agent's TLS and access control keys $(\crt{A}, \PAC{A})$ and a signed one-time key $\OTK{A}{i}, \signature{\OTK{}{i}}{\texttt{U1}}$. Subsequently, the \provider decrements the counter $\otkcounter{\aid{A}}{\aid{B}}$ by one.
    \label{step:communication_access_inquiry}
    \item \textbf{Receiving agent information verification:}  
    \B first verifies \A's  user's certificate $\crt{U1}$ including the user's public key $\PK{U1}$. \B also verifies the signatures on agent's \A information and the received $\OTK{}{}$ as follows:\label{agent_comm_sig_verify}  
    \begin{align*}
    \text{Verify}_{\PK{\texttt{U1}}}(& 
        \langle \aid{\A}, \ED{\A}, \PK{\A}, \PAC{\A}, \PK{Prov} \rangle, \signature{\A}{\U}) 
    \end{align*}
    \begin{align*}
    \text{Verify}_{\PK{\texttt{U1}}}(\langle \aid{A},\OTK{A}{i}, \rangle,  \signature{\OTK{}{i}}{\U})
    \end{align*}
    \item \textbf{Establishing a TLS connection between agents:}  
    \B initiates a TLS connection with \A, and both agents verify each other's certificates ($\crt{A}, \crt{B}$).\label{mtls_conn_agents} 

    \item \textbf{Token request sent:}
    \B sends \A its information and a signature $\signature{\B}{Prov}$ from the \provider (generated during agent registration: step \ref{step:agent_registration_ack}), along with one-time key $\OTK{A}{i}$, requesting access.

   \item \textbf{Token request received:}  
    \A verifies \texttt{U2}'s certificate, as well as $\signature{\B}{Prov}$:
    \begin{align}
    \text{Verify}_{\PK{Prov}}(\langle \aid{B}, \crt{B}, \ED{B}, \PAC{B}, \signature{\B}{\texttt{U2}} \rangle, \signature{\B}{Prov}) \nonumber
    \end{align}

     If $\OTK{A}{i}$ is valid, both agents perform a Diffie-Hellman (DH) key exchange protocol to derive a shared key:
     $$
        \myDH{A}{} = DH(\SOTK{A}{i}, \PAC{B}), \quad
         \myDH{B}{} = DH(\SAC{B}, \OTK{A}{i}) 
     $$
     $$
         \SDHK = KDF(\myDH{A}{}) = KDF(\myDH{B}{})
     $$

    \item \textbf{Token generation:}  
    \A creates the access token with a randomly generated nonce ($\getnonce$), issue ($\timestamp{issued}$) and expiration ($\timestamp{expire}$) timestamps, the maximum number of requests linked to this token ($\quota{max}$), and \B's access control key ($\PAC{B}$):
    $$
    \token = \text{Enc}_{\SDHK}(\langle \nonce, \timestamp{issued}, \timestamp{expire}, \quota{max}, \PAC{B} \rangle).
    $$
    \A stores the token and sends it to \B. 

    \item \textbf{Inter-Agent communication:}  
    \B receives the token and initiates the conversation to complete its task. For each subsequent request to \A, \B attaches the token. Upon receiving a request, \A verifies that the token was issued for \B (not for another agent), has not expired, and has not exceeded its usage quota. 
    Once a task is deemed completed, the token is discarded by both parties.

    \label{agent_comm_convo_init}
\end{enumerate}

\shortsection{Token reuse}
Once an \agent obtains a token, it can reuse it as long as it remains valid and hasn't exceeded its request limit. If \B holds a valid token for \A, it can skip ahead to step \ref{agent_comm_convo_init} to initiate secure communication. If the agents' existing TLS session is reset, the agents will establish a new TLS session (step \ref{step:agent_comm_tls_establish}) and proceed directly to agent communication (step \ref{agent_comm_convo_init}). The expiration time and request limit in the token offer a balance between security and performance considerations. A larger number of requests $\quota{max}$ reduces the overhead of contacting the \provider to obtain $\OTK{}{}$s, but increases the potential exposure in case an agent is compromised.

\subsection{Formal Protocol Analysis}
We formalized \saga using the state-of-the-art symbolic cryptographic analysis tool, \textsc{ProVerif}, to reason about cryptographic attackers \cite{Proverif}. Our \textsc{ProVerif} model precisely captures the \saga protocol description, using standard \textsc{ProVerif} protocol modeling techniques and assumptions. Within our \textsc{ProVerif} model of \saga, we encoded formal properties specifying the secrecy of the \saga token, authentication of communication between agents and the provider, and authentication of communication between any two agents. Using \textsc{ProVerif}'s automated reasoning capabilities, we automatically proved each of the afformentioned properties with respect to an attacker that can observe, intercept, modify, replay, reorder, and synthesize arbitrary messages on the network. This directly corresponds to the Dolev-Yao symbolic cryptographic protocol model \cite{Proverif}. For details on formal verification, see \Cref{sec:formal_details}.

\section{\revision{Extensions to the \saga architecture}}

\revision{We discuss several extensions to \saga to enable fault tolerance, scalability and resilience to server compromise in Section~\ref{sec:provider_architecture}, and describe the integration with the A2A protocol in Section~\ref{sec:a2a_integration}.}

\subsection{Provider Architecture Design} 
\label{sec:provider_architecture}
In our proposed architecture, the \provider is trusted, a model similar to that in existing systems as \textit{Active Directory} \revision{\cite{desmond2008active}} (based on Kerberos). We discuss below how the design can be augmented to  enable fault-tolerance, resilience to attacks and server compromises, and scalability.

\textbf{Fault-tolerance}. The centralized \provider design we propose can be made fault-tolerant by using standard distributed systems techniques. Specifically, the \provider functionality can be implemented as a RAFT~\cite{raft} or Paxos~\cite{paxos} cluster, where the agent registry is replicated across the RAFT nodes. Operations such as agent registration, policy update, and \revision{$\OTK{}{}$} retrieval are submitted to the leader of the RAFT cluster. Such systems are provisioned to tolerate a certain number of failures, for example typical deployments consist of 5 servers tolerating 2 faulty servers. We present fault tolerance results in Section~\ref{sec:fault_tolerance_scalability}.

\textbf{Protection against denial-of-service (DoS)}. Additionally, protection against DoS can be achieved by using rate limiters that control the volume and frequency of incoming requests, preventing attackers from overwhelming server resources with excessive traffic. Advanced rate limiters can detect abnormal traffic patterns, adapt their restrictions based on current server load, and implement tiered responses like introducing delays rather than complete blocking. While particularly effective against single-source attacks, rate limiters serve as a crucial first line of defense that maintains system availability during attacks, though they work best when combined with other protective measures like firewalls and DDoS protection services for comprehensive security against sophisticated distributed attacks.

\textbf{Resilience to compromised servers}. A compromised \provider might refuse agent registration, refuse to forward agent metadata, register malicious agents, or not follow agent policies. The \provider can be made resilient to these types of attacks by implementing it as a Byzantine-resilient service using existing protocols such as PBFT~\cite{pbft}.  In this design, a quorum of servers is required to participate in each operation ensuring that a majority of honest servers make the decision. Such a  system can tolerate $f$ compromised servers out of $3f+1$ participating servers. An alternative, more lightweight design is using fault-tolerant  algorithms (RAFT or Paxos), augmented with proactive auditing mechanisms to detect server compromises.

\textbf{Scalability.} Finally, the scalability of the system can be elastically increased by using a standard technique in databases~\cite{bigtable, spanner, megastore} called \textit{sharding}~\cite{silberschatz2011database, solat2024sharding}. In this approach the agent registry is partitioned across several entities called sharders, and we can partition based on the space of agent ID. Each sharder can be made fault-tolerant with a RAFT cluster, and a set of load-balancers forwarding the requests for different agent entries to the corresponding sharder. We present scalability results in Section~\ref{sec:fault_tolerance_scalability}.

\textbf{Federation.} The \provider service can be further decentralized by using federation, where several organizations each managing their own provider, participate in the protocol. Similar to cross-realm Kerberos \cite{rfc4120}, multiple providers establish trust relationships via shared cryptographic keys or certificates, enabling cross-domain agent authentication where agents from one provider can securely communicate with agents from other  providers.  This model allows organizations to maintain control over their agent registries and policies while enabling inter-organizational communication, with each provider validating its local agents during cross-boundary interactions.

\subsection{Integration with the A2A Protocol}
\label{sec:a2a_integration}

\revision{We discuss the integration of \saga with Google’s Agent2Agent (A2A) protocol~\cite{Surapaneni2025A2A}. A2A defines an agentic framework where AI agents
advertise public metadata and establish  secure communication with other agents.
While A2A defines a unified interface for structured task exchange via ``agent cards'', it lacks support for authentication, access control, and agent governance. Our integration bridges this gap by protecting agent cards with SAGA access control policies and by encapsulating A2A messages within SAGA’s secure communication layer. We highlight below the changes that enable integration with \saga.}

\revision{The \provider stores agent cards in the agent registry as part of each agent’s metadata entry  as per Step \ref{step:agent_registration_ack} of Agent Registration in \Cref{sec:agent_registration_by_user}.
\begin{align*}
     \agentregistry[\aid{A}] = \langle \uid{U}, \meta{A}, \acp{A}, \texttt{\textbf{A2ACard}}_{\texttt{A}}, \signature{\A}{\U}, \signature{\OTK{}{i}}{\U} \rangle
\end{align*}
Unlike A2A’s recommended deployment via public URLs, \saga protects these cards under user-specified access control policies. This ensures that only authorized agents may retrieve the agent cards.
Additionally, the integrity of the agent cards are protected with user signatures (Step \ref{step:agent_registration_genkey}, \Cref{sec:agent_registration_by_user}).}

\resizebox{\columnwidth}{!}{%
\begin{minipage}{\columnwidth}
\revision{\begin{align*}
    \signature{\A}{\U} = \signing{\SK{U}}{
    \langle \aid{A}, \ED{A}, \PK{A}, \PAC{A}, \PK{Prov}, \texttt{\textbf{A2ACard}}_{\texttt{A}} \rangle}
\end{align*}}
\end{minipage}
}
\revision{For inter-agent communication, the standard message in the \textit{msg} field is wrapped into an \texttt{A2A Request} during Step 8 of \Cref{sec:agent_communication}. 
\begin{align*}
      \langle \token, \texttt{msg} \rangle \rightarrow \langle \token, \texttt{\textbf{A2ARequest}(\texttt{msg})} \rangle
\end{align*}
In addition to the message content, \texttt{A2A Requests} include internal metadata such as task and message IDs, and the content type (e.g., text, image). Before forwarding the request to the agent's A2A stack, the \saga stack verifies the token for authenticity, freshness, and contact authorization as per Step \ref{agent_comm_convo_init} of \Cref{sec:agent_communication}. If any check fails, the message is discarded and never reaches the A2A layer, effectively preventing unauthorized task execution.}

\revision{\saga integrates with A2A through minimal changes that preserve the protocol's functionality, while augmenting its security.
The integration remains agnostic to agent design and task semantics, enabling secure, interoperable agent communication controlled by user policies.}

\section{Evaluation}

\revision{To evaluate \saga, we implement the full protocol (\ref{sec:implementation}), measure its overhead in \ref{sec:protocol_overhead}, and evaluate it on three  agentic tasks in \ref{sec:task_completion}. We measure the cost of fault tolerance and protocol scalability in \ref{sec:fault_tolerance_scalability}. In Appendix \ref{sec:attacker_evaluation} we implement 8 attackers within the capabilities defined in the threat model (\ref{sec:threat_model}) and evaluate \saga's resilience.}

\subsection{Implementation}
\label{sec:implementation}

The \provider is implemented as an HTTPS service.
Inter-agent communication is conducted over TLS configured with mutual authentication, with protocol-level authentication and encryption enforced via ephemeral session keys.
Our framework is agnostic to the underlying LLM-agent implementation. This design enables seamless integration with arbitrary agent implementations. \revision{We also implemented the integration of \saga with A2A, described in~\ref{sec:a2a_integration}.}

All cryptographic operations in the protocol are built on Curve25519 \cite{bernstein_25519}. Both long-term and ephemeral keys are generated using the \texttt{X25519} elliptic-curve Diffie-Hellman (ECDH) \cite{rfc7748} scheme, using 256-bit shared secrets. Certificates adhere to the X.509 PKI standard \cite{rfc5280} and are issued by an internal certificate authority (\ca) deployed as part of the provider. All digital signatures and key derivation steps utilize the \texttt{SHA256} hash function \cite{dworkin2015sha}.

For LLM agents, we experimented with a local Qwen-2.5 \cite{qwen2.5} 72B model running on NVIDIA H100, as well as two OpenAI models hosted in the cloud and accessed via API.

\subsection{Overhead Evaluation}
\label{sec:protocol_overhead}

\shortsection{Cryptographic Overhead}
We begin with measuring the cryptographic overhead of core protocol operations at the user, \provider, and agents.
These costs cover cryptographic primitives such as hashing, key generation, signing, verification, and Diffie-Hellman key exchange. As shown in \Cref{tab:overhead} (Appendix \ref{apdx:more_crypto}) most operations are lightweight, on the order of several ms.

\shortsection{\revision{Key Management Overhead}} 
\revision{\saga relies on three classes of cryptographic keys, each with distinct lifetimes and rotation patterns: short-term one-time keys ($\OTK{}{}$), medium-term access control keys $(\PAC{}, \SAC{})$, and long-term identity keys $(\PK{}, \SK{})$. $\OTK{}{}$s are ephemeral and rotated frequently once consumed for the derivation of access control tokens. On the other hand, access control keys are medium-term, typically rotated on a weekly or biweekly basis to balance security with operational stability. Long-term identity keys are rotated infrequently, commonly every 30 to 90 days, following established key management guidelines \cite{NIST80057r5}.}

\begin{figure}[h]
    \centering
    \includegraphics[width=\linewidth]{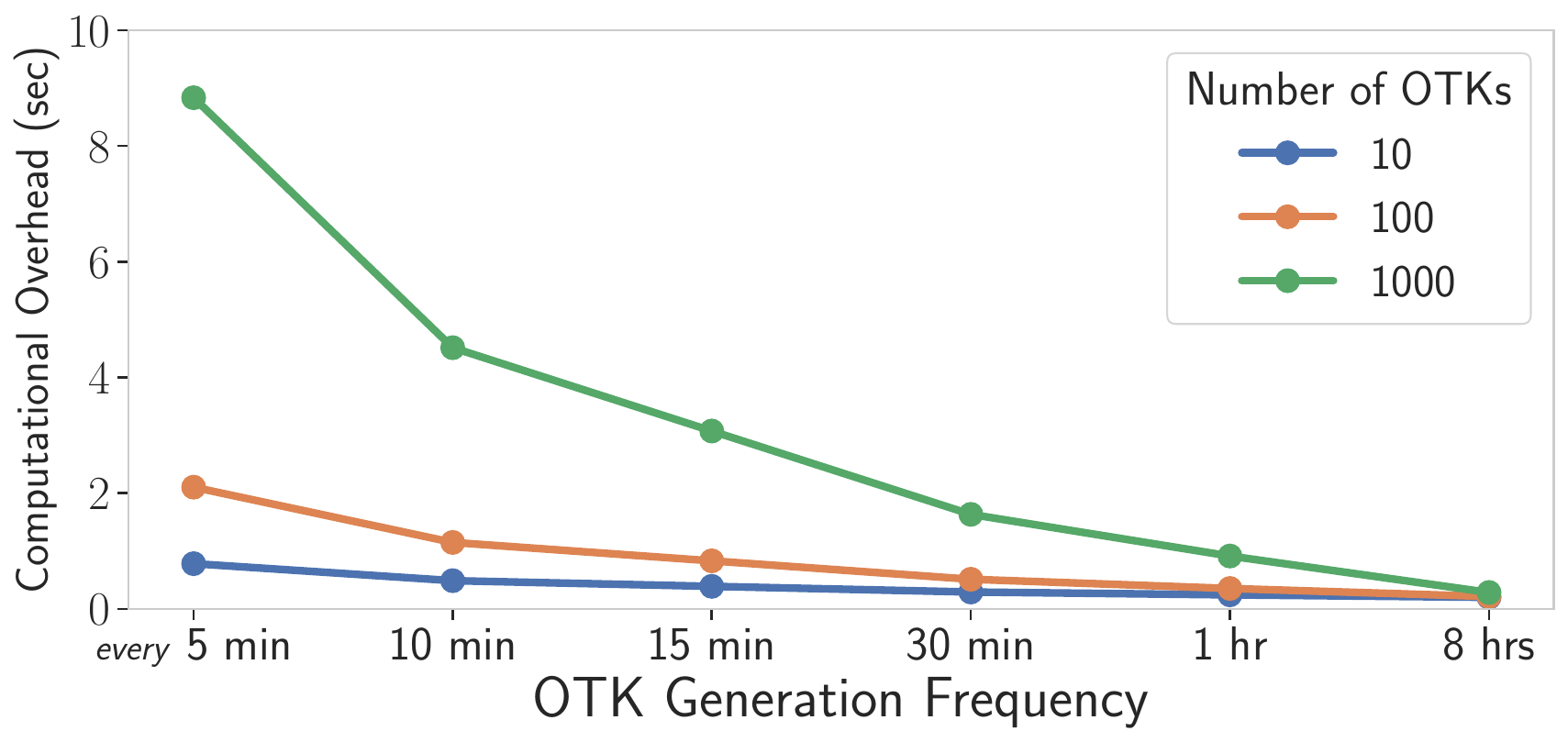}
    \caption{\revision{Computational overhead of $\OTK{}{}$ generation for the user, as a function of  frequency and key-chain length. Even with frequent generation (every 5 minutes) and long key-chains (1000 $\OTK{}{}$s), the total cost remains low.}}
    \label{fig:comp_otk_replenish}
\end{figure}

\revision{\Cref{fig:comp_otk_replenish} measures the overhead for OTK generation at the user, showing that even under frequent generation (1000 $\OTK{}{}$s every 5 minutes over an 8-hour period) the total computational cost remains under 10 seconds for a single user. The computational cost at the \provider (validation and storage of the $\OTK{}{}$s) does not exceed 0.5 seconds under the same conditions (1000 $\OTK{}{}$s/5min). In contrast, both access control and identity key pairs require only $0.11$ milliseconds to generate on commodity hardware, and due to their infrequent rotation, their cost is effectively amortized.}

\revision{\Cref{fig:comp_act_gen_freq} reports the total cost of deriving access control tokens with different configurable lifetimes ($L$) from $\OTK{}{}$s for a single initiating agent with 1, 10 and 100 receiving agents. The process includes a Diffie-Hellman handshake, the encryption and decryption (validation) of the token as described in \Cref{sec:agent_communication}. 
Even for very short lifetimes (1 minute), the total cost of derivation of 144K tokens is just below 400 seconds over the span of 1 day for a single agent.}

\revision{We conclude that a longer lifetime reduces reliance on the \provider but increases the window for compromised agents to operate without interruption. Shorter lifetimes mitigate security risks by requiring more frequent cryptographic validation, at the expense of additional overhead.}

\begin{figure}[h]
    \centering
    \includegraphics[width=\linewidth]{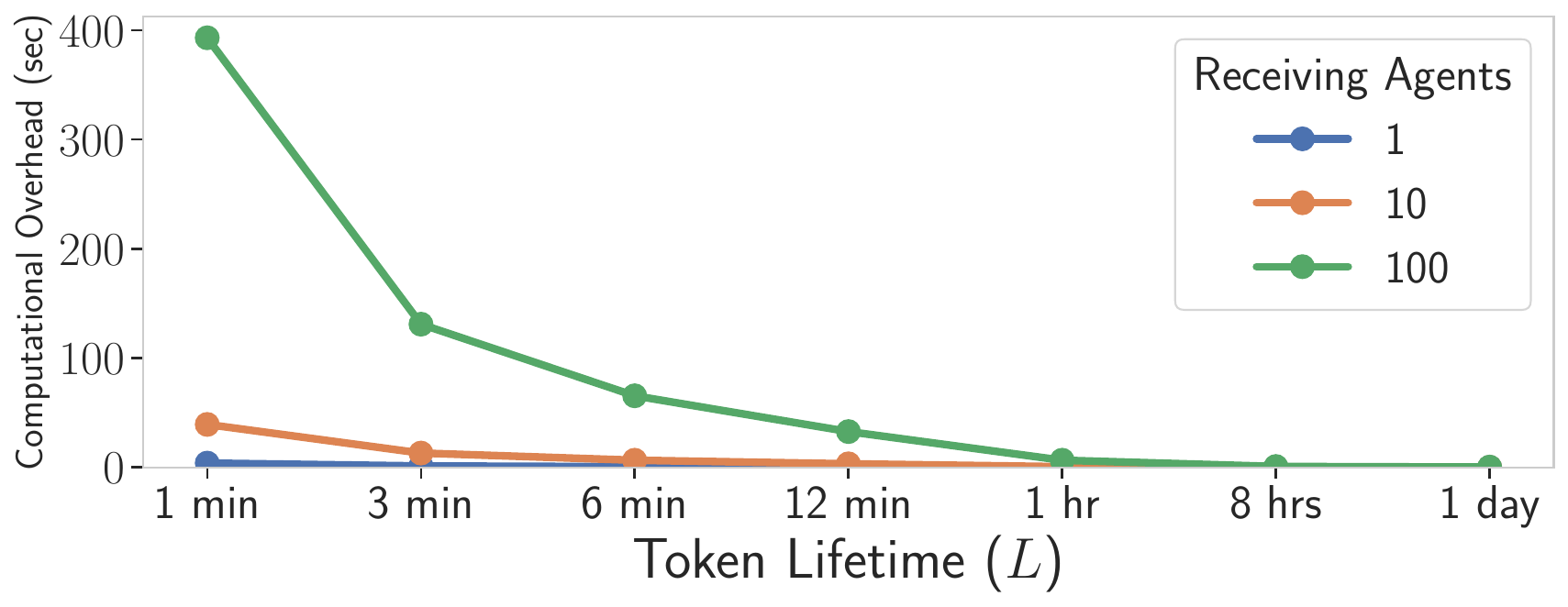}
    \caption{\revision{Computational overhead of access control token derivation between one initiating agent and 1, 10 and 100 receiving agents, varying by token lifetime ($L$). Even for short lifetimes (1 minute), the total cost remains low.}}
    \label{fig:comp_act_gen_freq}
\end{figure}

\revision{Importantly, all aforementioned overheads are very small compared to the total task execution times (\Cref{sec:task_completion}), which typically last at least a few minutes, depending on task complexity and LLM backend latency.} %

\shortsection{Protocol Overhead}
We measure the overhead introduced by \saga's access control and provider coordination mechanisms (\Cref{sec:protocol_overhead}). Specifically, we analyze the overhead incurred by an initiating agent \B issuing \( m \) requests to a receiving agent \A. This includes a network component for establishing secure communication, and a cryptographic component $t_{\mathrm{crypto}}$ for certificate validation, signature verification, Diffie-Hellman key exchange, key derivation, token encoding, and symmetric encryption.
The total protocol overhead is modeled as:
\begin{align}
    c_{\text{proto}}(m) = \left( \mathrm{RTT}_{B,P} + t_{\mathrm{crypto}} \right) \cdot \left\lceil \frac{m}{\quota{max}} \right\rceil,
\end{align}
where $P$ is the \provider, and $\mathrm{RTT}_{B,P}$ is the round-trip time for agent \B contacting the \provider and receiving a response.
Each authorization cycle involves agent \B retrieving metadata and a one-time key for agent \A from the \provider. This round-trip, along with local cryptographic operations, must be performed once every $\quota{max}$ requests, as the token quota is exhausted.

\begin{figure}[h]
    \centering
    \includegraphics[width=\linewidth]{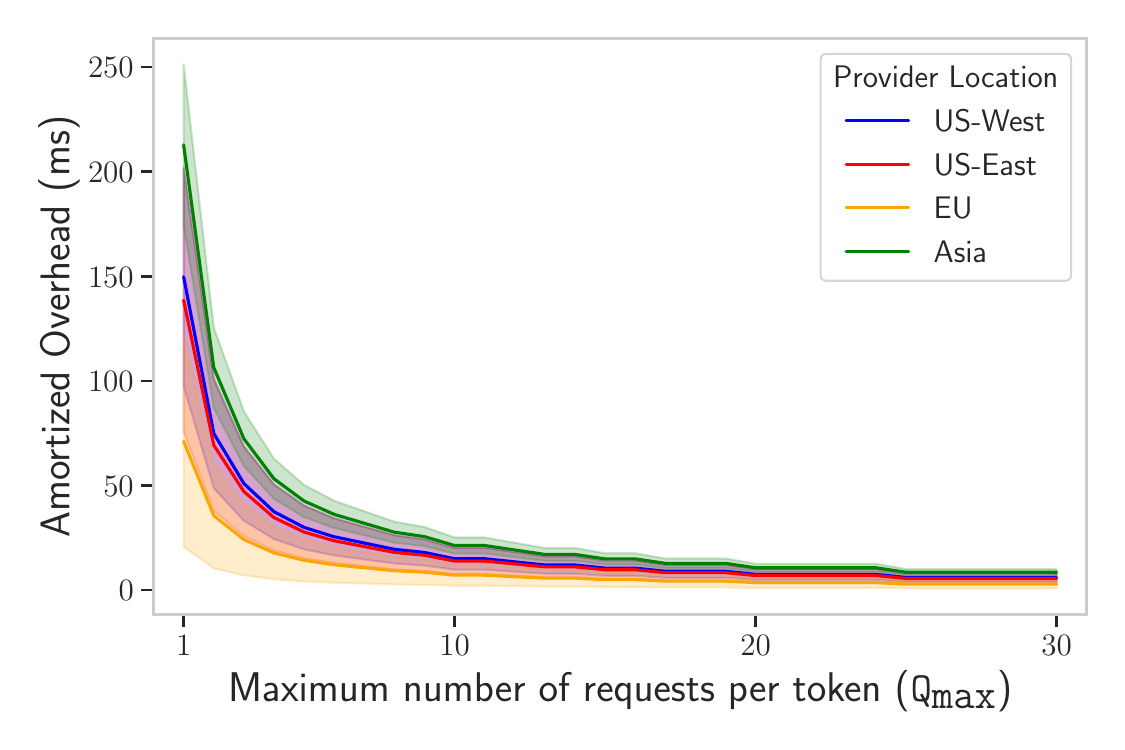}
    \caption{Amortized protocol overhead per request $\bar{c}_{\text{proto}}(m)$ as a function of maximum number of requests token is reused. We measure the overhead for several geographic locations for the \provider. The shaded region reflects variability for agents position worldwide.}
    \label{fig:provider-fixed}
\end{figure}

\begin{figure}[h]
    \centering
    \includegraphics[width=\linewidth]{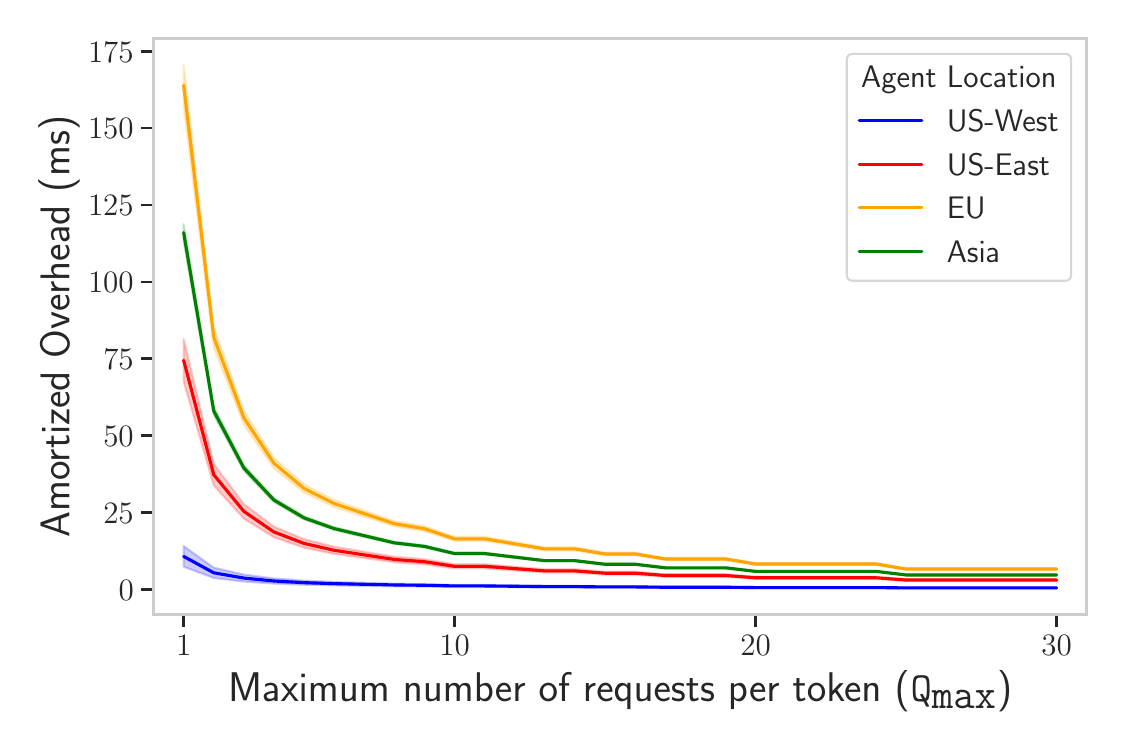}
    \caption{Amortized protocol overhead per request $\bar{c}_{\text{proto}}(m)$ as a function of maximum number of requests token is reused. We measure the overhead for several geographic locations for the initiating agent, where the \provider is fixed in US-West. The shaded region captures variability under sampled network conditions.}
    \label{fig:agent-fixed}
\end{figure}

We sample round-trip times ($\mathrm{RTT}_{B,P}$) from empirical measurement distributions using monitors in US-East, US-West, Europe and Asia, made available by CAIDA \cite{caida_rtt} and AWS \cite{aws_cloudpingco}, and use these to approximate protocol overhead.
\Cref{fig:provider-fixed} shows the amortized protocol setup overhead:
$$
\bar{c}_{\text{proto}}(m) = \frac{c_{\text{proto}}(m)}{m}
$$
as a function of token quota \( \quota{max} \), using \( m = 100 \) requests, where the measured cryptographic overhead \( t_{\mathrm{crypto}} = 7\,\mathrm{ms} \).
As \CR{shown}, the protocol overhead decreases sharply with increasing token quota $\quota{max}$, demonstrating how it can be effectively amortized across inter-agent interactions. Moreover, the overhead is low for all \provider geolocations, with slightly higher overhead for Europe and Asia. 

We also evaluate the effect of geolocation on protocol overhead by varying the placement of the initiating agent under a fixed \provider location (US-West).
The overhead is very low—under 25ms when agents interact for at least 4–5 requests (\Cref{fig:agent-fixed}). This effect is even more pronounced when the agent and \provider are geographically close.

\subsection{Task Completion}
\label{sec:task_completion}
\saga operates as a protocol layer above the underlying LLM agents, whose communications are not impacted by the protocol.
To illustrate \saga in realistic agent interactions, we deploy three types of agents: (a) \textit{Calendar} agents that determine a mutually available time and schedule a meeting, (b) \textit{Email} agents that extract relevant expense-related emails and  collaborate to submit an expense report, and (c) \textit{Writing} agents with different expertise areas collaborating to write a blogpost. As expected, all agents successfully complete their tasks.  
See \Cref{app:tasks} for more details.

\begin{table}[ht]
\centering
\begin{tabular}{llcc|c}
\toprule
\multirow{2}{*}{\textbf{Task}} & \multirow{2}{*}{\textbf{LLM Backend}} & \multicolumn{2}{c}{\textbf{Standard Cost}} & \textbf{\saga} \\
& & LLM & Networking & \textbf{Overhead}\\
\midrule
Calendar & \texttt{GPT-4.1-mini} & 50.001 & 0.791 & 0.165 \\
Email & \texttt{GPT-4.1} & 26.862 & 1.319 & 0.165 \\
Writing & \texttt{Qwen-2.5} & 363.563 & 1.319 & 0.165 \\
\bottomrule
\end{tabular}
\caption{Task execution time (in seconds).  \A, \B, and the \provider are located in Asia, Europe, and US-West, respectively, and the token quota is 10. \textbf{Standard Cost} is the minimum runtime for two agents communicating directly without \saga, including LLM cost and network latency.}
\label{tab:llm-latency-breakdown}
\end{table}

We measure the standard task completion cost \ie the time taken by the LLM to generate responses, and the network latency (\Cref{tab:llm-latency-breakdown}, under \textbf{Standard Cost}).
The LLM response time depends on both model execution speed and task complexity. For example, using a local \texttt{Qwen-2.5} model instead of the cloud-based \texttt{GPT-4.1} model for the Email task increases the runtime from 26.862 to 43.730 seconds, as the Qwen-2.5 (72B) model is slower than the highly optimized models served by OpenAI. Tasks like blog post writing require substantially more input and output tokens and result in longer runtimes, as observed for the Writing task.
Since most of the task completion time is spent by LLM-agents during intermediate planning \cite{yao2023react}, and tool calls \cite{schick2023toolformer}, the amortized overhead of our protocol is significantly lower in comparison. For example, even when agents and the \provider are geographically distant, the protocol overhead accounts for less than $0.6\%$ of the end-to-end cost of completing the fastest calendar task.

\subsection{\revision{Fault Tolerance and Scalability}}
\label{sec:fault_tolerance_scalability}

\begin{figure*}[htp]
     \centering
     \begin{subfigure}[b]{0.32923\textwidth}
         \centering
         \includegraphics[width=\textwidth]{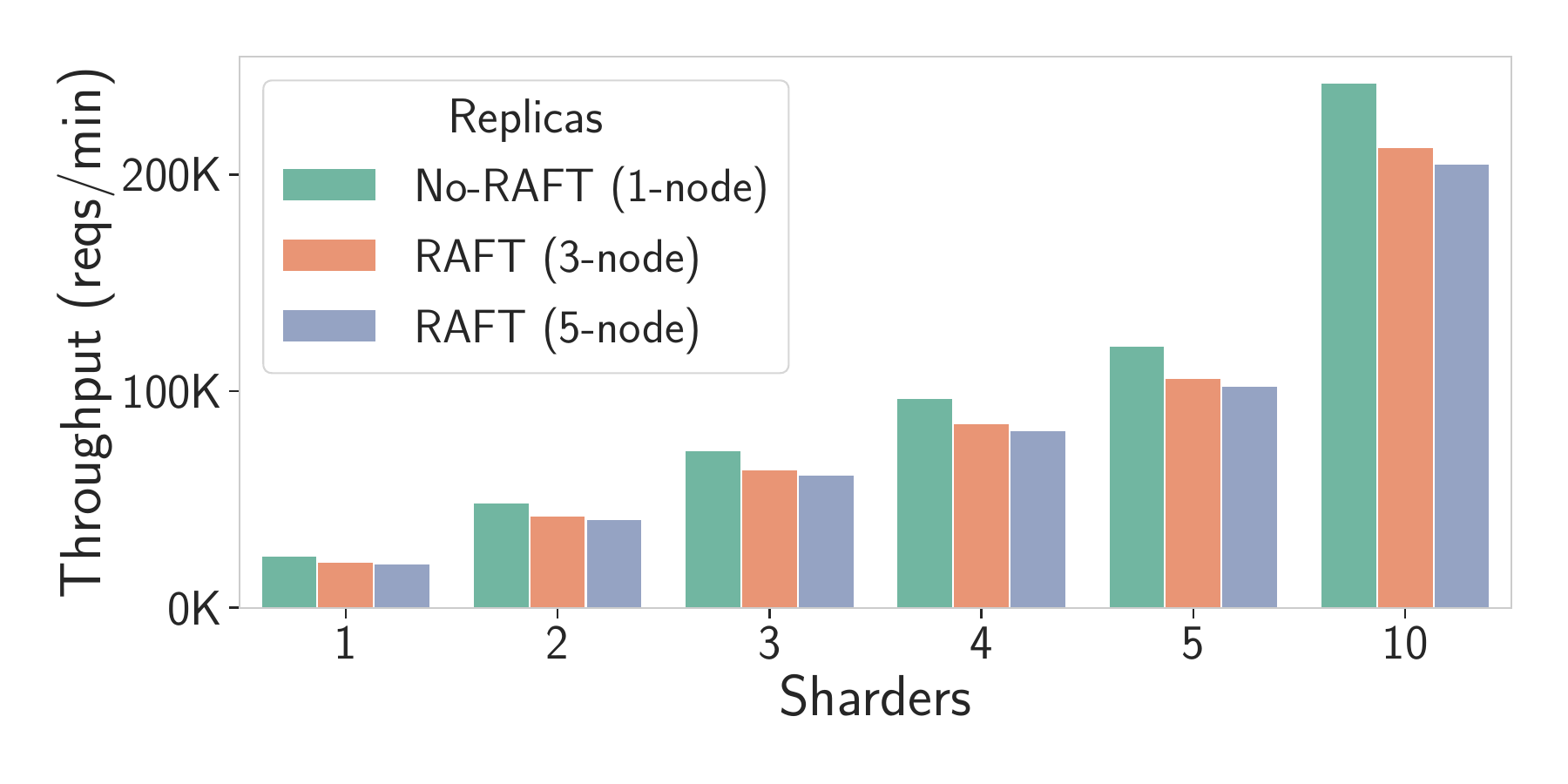}
         \caption{\revision{$\OTK{}{}$ Request Throughput}}
         \label{fig:otk_req}
     \end{subfigure}
     \hfill
     \begin{subfigure}[b]{0.32923\textwidth}
         \centering
         \includegraphics[width=\textwidth]{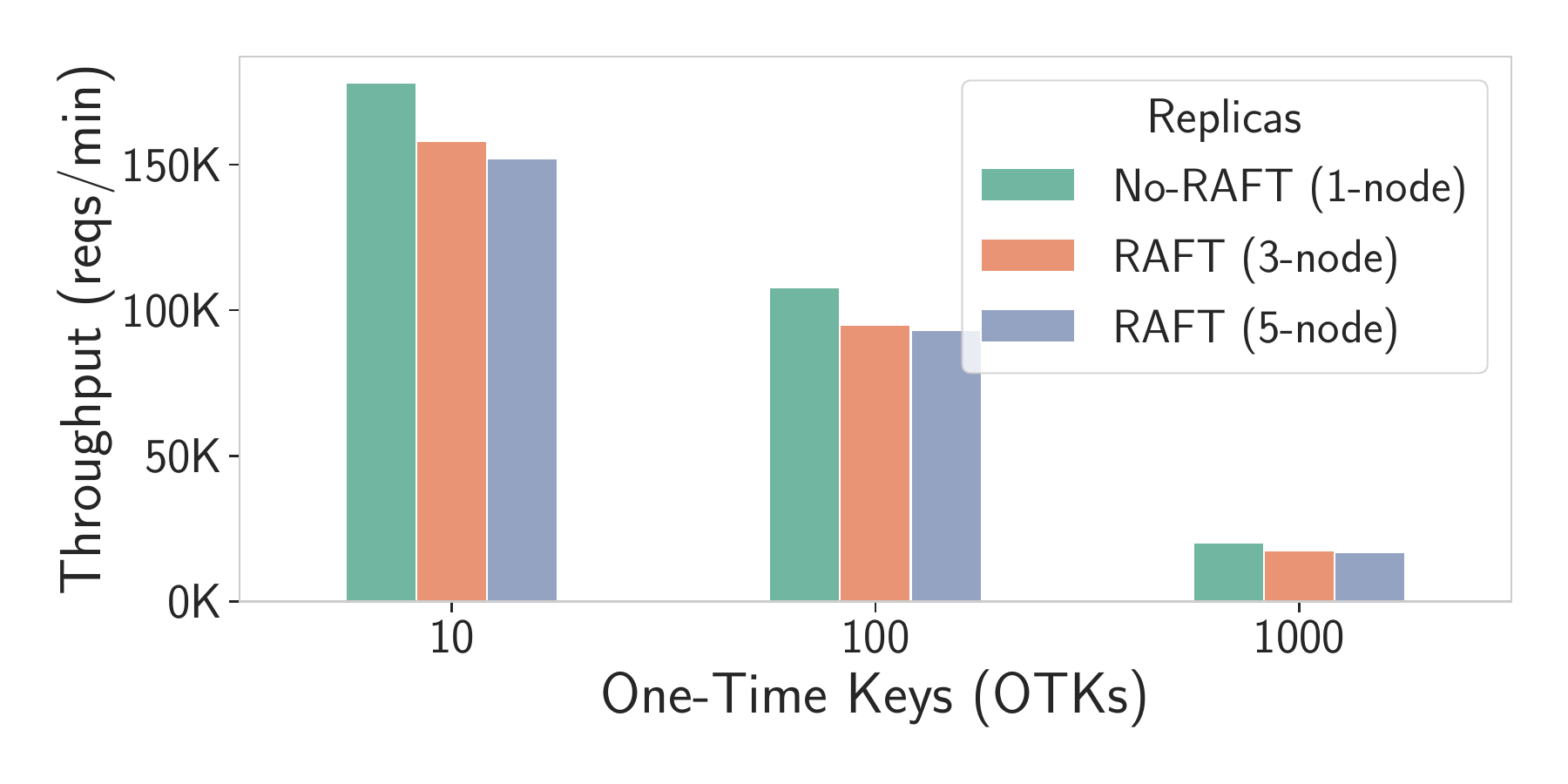}
         \caption{\revision{$\OTK{}{}$ Refresh Throughput %
         }}
         \label{fig:otk_rep}
     \end{subfigure}
     \hfill
     \begin{subfigure}[b]{0.32923\textwidth}
         \centering
         \includegraphics[width=\textwidth]{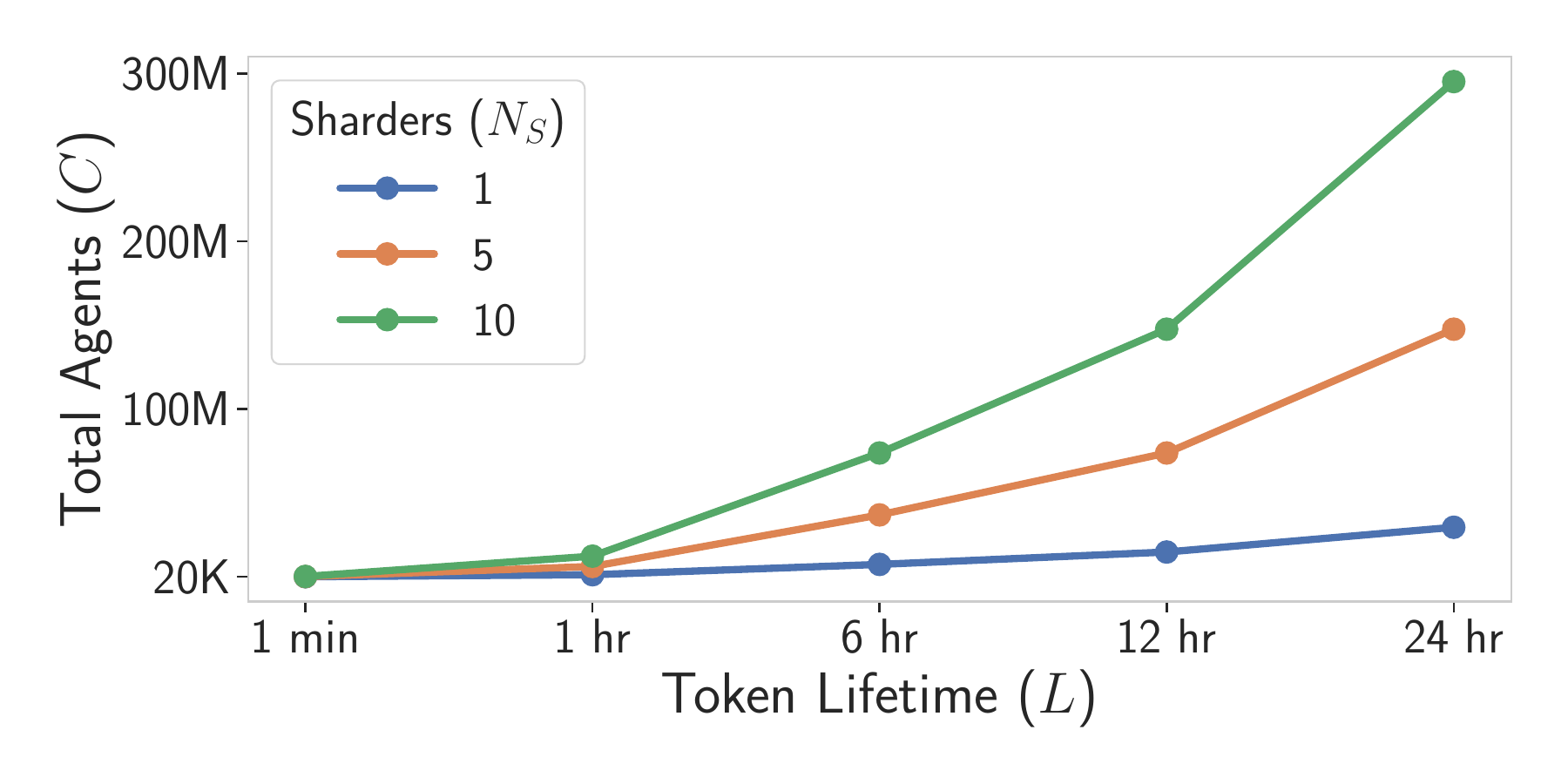}
         \caption{\revision{Total Capacity $C$ of the system.}}
         \label{fig:agent_capacity}
     \end{subfigure}
        \caption{\revision{SAGA's Throughput and Capacity for different protocol operations.}}
        \label{fig:ft_scale}
\end{figure*}

\revision{We evaluate \saga's scalability and fault tolerance under varying deployment configurations. Specifically, we measure the throughput of the \provider for core operations; agent registration, $\OTK{}{}$ request (i.e., issuance of one-time keys to agents), and $\OTK{}{}$ refresh (i.e., generation of new OTKs by the user). We vary system parameters such as replication factor (RAFT nodes), number of sharding workers $N_S$, $\OTK{}{}$ key-chain length, and access control \token lifetime. Our findings show that \saga maintains high throughput under replication, scales linearly with added compute, and supports large agent populations through configurable token lifetimes.}

\shortsection{\revision{Setup}}
\revision{We deploy a \provider backed by a replicated RethinkDB cluster. RethinkDB~\cite{rethinkdb} is a distributed, open-source database using the RAFT consensus algorithm for strong consistency and fault tolerance. For replication cost evaluation, we consider typical cluster configurations with 3 and 5 RAFT nodes (supporting 1 and 2 crash faults), and as baseline a configuration with only 1 node (no fault tolerance). For scalability evaluation, we vary the number of \textit{sharders} and observe throughput assuming that each sharder (consisting of a RAFT cluster) runs on a separate machine and requests are routed by a proxy to the right sharder. To analyze the system's sensitivity to storage-related operations, we vary the number of submitted $\OTK{}{}$s between 10, 100, and 1000 keys. Experiments are conducted on a workstation with a 16-core AMD Threadripper PRO 5955WX CPU and Samsung's MZ1L21T9HCLS-00A07 SSD.}

\shortsection{\revision{Fault Tolerance}} \revision{Figures~\ref{fig:otk_req}, \ref{fig:otk_rep}, and \ref{fig:reg_agent_vary_otks} demonstrate that making the \provider fault-tolerant (3, 5 RAFT nodes) introduces negligible throughput degradation across key operations. For example, for $\OTK{}{}$ Requests (Figure~\ref{fig:otk_req}), the No-RAFT (1-node) setup achieves 242K requests per minute, compared to 212K for 3-node and 204K for 5-node RAFT configurations, which translates to a throughput decrease of 12-15\%. In Figure~\ref{fig:otk_rep}, the cost of replication for $\OTK{}{}$ refresh is lower, dropping from 173K (No-RAFT) to 153K (3-node RAFT), a difference of just 20K requests per minute ($\sim$11\%). For more I/O-intensive operations such as refreshing large key-chains (1000 $\OTK{}{}$s), the tradeoff shrinks further to only 2K requests per minute. 
Agent registration performance remains largely stable even under 5-node replication, regardless of the number of $\OTK{}{}$s provisioned per agent achieving 511K requests per minute. Similarly, $\OTK{}{}$ request and refresh incur minimal overhead.}

\shortsection{\revision{Scalability}} \revision{As expected, throughput scales linearly with the number of sharders. As shown in Figures \ref{fig:otk_req}, \ref{fig:reg_agent_vary_otks} and \ref{fig:replenish_vary_otks} adding just 10 sharders increases $\OTK{}{}$ throughput nearly tenfold, indicating no early saturation and achieving 178K to 511K requests per minute depending on the operation. We omit the latency of forwarding the request to the right sharder, as within the same datacenter this would be very small. Experiments on AWS including the cloud latency show a similar trend (see \Cref{apdx:aws}). We expect that in production, \provider can scale to hundreds of sharders.}

\shortsection{\revision{Agent Capacity}}
\revision{We seek to  answer the question: \textit{How many active agents can the system support at any given time?} We define this as the number of agents a \provider can \underline{continuously} serve with $\OTK{}{}$s.
Let $L = \texttt{T}_{\texttt{expire}} - \texttt{T}_{\texttt{issue}}$ denote the \token lifetime. Each agent needs to contact the \provider only once per $L$ to obtain an $\OTK{}{}$, which it then uses to generate a token and communicate with another agent throughout the interval. Hence, the provider’s total supported population is:
$C = \mathcal{T}(N_S) \cdot L$
where $\mathcal{T}(N_S)$ is the \provider’s $\OTK{}{}$ issuance throughput with $N_S$ sharder nodes. 
As shown in \Cref{fig:agent_capacity}, with 10 sharders and 24-hour tokens, the system can support up to 300 million agents. These results demonstrate \saga's ability to operate under realistic, large-scale deployment scenarios.}

\shortsection{\revision{Sensitivity to the $\OTK{}{}$ Key-Chain Length}}
\revision{We analyze the effect of increasing the $\OTK{}{}$ key-chain length (number of  $\OTK{}{}$s submitted to the \provider) on the throughput for agent registration and $\OTK{}{}$ refresh. 
We omit $\OTK{}{}$ request, as each request consumes exactly one $\OTK{}{}$ by design. As expected, Figures~\ref{fig:reg_agent_vary_otks} and \ref{fig:otk_rep} show a gradual throughput decrease as the number of submitted $\OTK{}{}$s increases. This is due to the increased payload size and the cost of writing more state to the replicated agent registry.} 

\revision{While longer $\OTK{}{}$ key-chains incur higher cost, \saga maintains high throughput across practical configurations. For example, agent registration with 1000 $\OTK{}{}$s still exceeds 40K requests per minute with 10 sharders (Figure~\ref{fig:reg_agent_1000_otks}), and $\OTK{}{}$ refresh remains comparably efficient across RAFT configurations (Figure~\ref{fig:otk_rep}). Despite processing the same number of $\OTK{}{}$s, registration outperforms $\OTK{}{}$ refresh due to differences in storage operations: registration involves appending to the registry whereas $\OTK{}{}$ refresh updates existing entries.
These results confirm that users can pre-provision large key batches with small performance penalties, enhancing resilience under intermittent connectivity.}

\section{Related Work}
\label{sec:related_work}

This section reviews existing frameworks for secure agent interactions and  current implementation landscapes, highlighting the gaps our work addresses. We also explore existing multi-agent attacks, highlighting their limited scope.

\shortsection{High-Level Designs for Inter-Agent Interaction}
Several frameworks and protocols have been proposed to govern AI agent interactions securely. South et al. \cite{south2025authenticated} introduce a framework for authenticated delegation using OAuth 2.0 design patterns, where users register their agents with providers and issue delegation tokens. %
Chan et al. \cite{chan2025infrastructure} outline agent infrastructure protocols defining interaction standards, 
focused on three core functions: attribution, interaction, and response, 
Chan et al. \cite{chan2024ids} propose a system for agent identification 
that tracks AI systems along with their context windows and initial users. %
Shavit et al. \cite{shavit2023practices} discuss governance practices for agentic systems, highlighting unique agent identification and interruptibility as critical features. %
None of these designs are actually implemented and evaluated in a real system.

\shortsection{LLM Agent Development Frameworks} \revision{LangChain \cite{langchainaiAgentProtocol}, AutoGen \cite{wu2024autogen}, and smolagents \cite{smolagents} are development libraries that help build LLM agents, but do not address governance—how users control their agents or how agents securely discover and communicate with each other. We used smolagents in our experiments, but SAGA is compatible with any other agent development framework. SAGA provides user-level control of agent interaction with other agents and secure agent communication.}

\shortsection{Inter-Agent Protocol Implementations}
Recent surveys \cite{tran2025multi} reveal that 
implementations typically assume pre-established connections between agents with static communication patterns. %
Smyth OS\footnote{\url{https://smythos.com/}} offers a portal for agent creation and integration limited within an particular 
organization. Current software implementations for agent interactions generally fall short of enabling secure cross-organizational communication. While frameworks like LangChain's Agent Protocol\footnote{\url{https://github.com/langchain-ai/agent-protocol}} provide specifications for interactions, they do not address critical aspects like authentication or access control mechanisms. Commercial platforms like Amazon Bedrock \cite{amazonCreatingAsynchronous} enable agent orchestration but assume all agents belong to the same restricted environment. Packages like AutoGen \cite{wu2024autogen} support multi-agent interactions, but only when agents are all running locally.
AACP \cite{royce2025ai} introduces a cryptographic foundation using verifiable credentials but adopts a fully peer-to-peer architecture without mediation layers to govern agent interactions.

\shortsection{Key Pre-Distribution and Trust Models}
\revision{Users of Signal and Matrix provide a server with signed one-time keys and ephemeral keys in-advance to allow offline shared key establishment; users either choose to trust the central server, or verify identity keys off-band in the case server compromise is a concern. \cite{Albrecht_Dowling_Jones, Kret_Schmidt}. SAGA employs a similar technique, allowing users to register their agents and provide one-time keys with a \provider to enable fine access control to agents, in the spirit of Kerberos \cite{rfc4120}.
Similar to Signal, users who do not trust the identity service (third party in \saga) must verify identity keys off-band. Additionally, the \provider can be made resilient to compromise, as discussed in Section \ref{sec:provider_architecture}.}

\shortsection{Attacks on Multi-Agent Systems}
Several works examine adversarial propagation in multi-agent communication \cite{gu2024agent, lee2024prompt, amayuelas2024multiagent, he2025red, yu2024infecting}, where rogue agents can propagate malicious outputs via interactions with other agents. Other works consider fixed communication patterns with slightly different goals, such as multi-agent debate \cite{amayuelas2024multiagent} and question-answer collaboration \cite{ju2024flooding}.
In orchestrated multi-agent systems, Triedman et al. \cite{triedman2025multi} describe attacks against orchestrator agents using adversarial content via metadata.
Khan et al. \cite{khan2025textit} introduce an attack approach for pragmatic multi-agent LLM systems operating under real-world constraints like token bandwidth limits and message latency. %
\saga provides protection against such attacks through its token-driven system, which explicitly limits the number of interactions between agents. Additionally, benign agents can easily update their contact policies to block malicious agents, preventing any widespread ``outbreak." %

\section{Conclusion}
\label{sec:conclusion}

\saga establishes a scalable framework for secure inter-agent communication that balances security, autonomy, and governance through a \provider-mediated architecture enforcing user policies. 
Unlike prior works that only offer conceptual designs or high-level architectures for agent governance \cite{south2025authenticated,chan2025infrastructure,chan2024ids,shavit2023practices}, \saga provides the first concrete protocol specification with strong formal security guarantees and a reference implementation. 
SAGA is compatible with existing agent protocols, such as A2A and Model Context Protocol \cite{modelcontextprotocolIntroductionModel} for standardized tool-use 
and it can be integrated with defenses against prompt-injection attacks \cite{debenedetti2025defeating}, and privacy-preserving data minimization techniques \cite{bagdasarian2024airgapagent}.

{\footnotesize
\bibliographystyle{IEEEtran}
\bibliography{references}
}

\appendix
\subsection{Protocol Figures}

Here we provide detailed protocol diagrams for \revision{user registration (\Cref{sec:user_registration}), } agent registration (\Cref{sec:agent_registration_by_user}) and agent communication (\Cref{sec:agent_communication}), under \revision{\Cref{fig:user_registration}, }\Cref{fig:agent_registration} and \Cref{fig:agent_communication} respectively.

\begin{figure}[t!]
  \centering
  \setmsckeyword{}
  \drawframe{no}    %
  \begin{msc}[
    /msc/title top distance=0cm,
    /msc/first level height=0cm,
    /msc/last level height=0.1cm,
    /msc/head height=0.2cm,
    /msc/instance width=0cm,
    /msc/head top distance=0cm,
    /msc/foot distance=-0.0cm,
    /msc/instance width=0cm,
    /msc/condition height=0.2cm
    ]{}
    \setlength{\instwidth}{0\mscunit} %
    \setlength{\instdist}{3cm}  %

    \declinst{U}{             
     \user
    }{}

    \declinst{P}{
      \provider
    }{} 

    \nextlevel[1]
    \action*{\parbox{134pt}{\footnotesize{$
        \text{Generate } \uid{U}, \texttt{passwd} \\
        \text{Generate } \SK{U}, \PK{U} \\
        \text{Get } \crt{U} = \createcert{\SK{CA}}{ \uid{U}, \PK{U}}  \\
        \text{Get \provider certificate and } \PK{Prov} 
    $}}}{U}

    \nextlevel[4]
    \condition{\messLabel{{\footnotesize Establish TLS connection}}}{P,U}

    \nextlevel[3]
    \mess{\messLabel{\text{$ \uid{U}, \texttt{passwd}, \crt{U} $}}}{U}{P}

    \nextlevel[1]
    \action*{\parbox{115pt}{\footnotesize{$
          \text{Verify user identity using service \emph{S}} \\
          \userregistry[\uid{U}] \text{=} \langle \texttt{H(passwd)}, \crt{U} \rangle
    $}}}{P}

    \nextlevel[3]
    \mess{\messLabel{\text{Confirmation}}}{P}{U}

    \nextlevel[1]
   \end{msc}

  \caption{\textbf{User Registration} protocol for user \U.}
  \label{fig:user_registration}
\end{figure}

\begin{figure*}%
  \centering
  \setmsckeyword{}
  \drawframe{no}    %
  \begin{msc}[
    /msc/title top distance=0cm,
    /msc/first level height=.1cm,
    /msc/last level height=0.4cm,
    /msc/head height=0cm,
    /msc/instance width=0cm,
    /msc/head top distance=0.5cm,
    /msc/foot distance=-0.0cm,
    /msc/instance width=0cm,
    /msc/condition height=0.2cm
    ]{}
    \setlength{\instwidth}{0\mscunit} %
    \setlength{\instdist}{10cm}  %

    \declinst{U}{
      \user
    }{}

    \declinst{P}{
      \provider
    }{}

    \action*{\parbox{165pt}{\footnotesize{$
            \text{Generate }  \aid{A}, \ED{A}, \acp{A} \\
             \text{Generate }  (\PK{A}, \SK{A}) \text{ and } (\PAC{A}, \SAC{A}) \\
             \text{Get }\crt{A} = \createcert{\SK{CA}}{\langle \aid{A}, \PK{A} \rangle} \\
             \text{Generate } (\OTK{A}{1}, \SOTK{A}{1}), \dots, (\OTK{A}{N}, \SOTK{A}{N}) \\
             \signature{\A}{\U} = \signing{\SK{U}}{
               \langle \aid{A}, \ED{A}, \PK{A}, \PAC{A}, \PK{Prov} \rangle
             } \\
             \signature{\OTK{A}{i}}{\U} = \signing{\SK{U}}{
               \OTK{A}{i}
             }
    $}}}{U}

    \nextlevel[5]
    \condition{\messLabel{{\footnotesize Establish TLS connection}}}{U,P}
    \nextlevel[3]
    \mess{\messLabel{\text{($\uid{U}$, \texttt{passwd})}}}{U}{P}

    \nextlevel[1]
    \action*{\parbox{75pt}{\footnotesize{$
            \text{Verify user credentials}. 
    $}}}{P}

    \nextlevel[3]
    \mess{\messLabel{\text{$(\aid{A}, \ED{A}, \acp{A})$, $\PK{U}$, $\crt{A}$, $\PAC{A}$, $\OTK{A}{i}$, $\signature{\OTK{}{i}}{\U}, i \in [1, N]$}}}{U}{P}

    \nextlevel[1]
    \action*{\parbox{185pt}{\footnotesize{$
            \text{Check uniqueness of } \aid{A} \text{ and } \ED{A} \\
            \text{Verify } \crt{A} 
            \text{ and signatures } \signature{\A}{\U}, \signature{\OTK{}{i}}{\U}, i \in [1, N] \\
            \text{Compute } \meta{A} = \{\ED{A}, \crt{A}, \PAC{A}, \OTK{A}{i}, i \in [1,N] \} \\
            \text{Store } \agentregistry[\aid{A}] = \langle \uid{U}, \meta{A}, \acp{A}, \signature{\A}{\U}, \signature{\OTK{}{i}}{\U} \rangle \\
            \signature{\A}{Prov} = \signing{\SK{Prov}}{\langle \crt{U}, \ED{A}, \PAC{A}, \signature{\A}{\U} \rangle}
    $}}}{P}

    \nextlevel[5]
    \mess{\messLabel{$\signature{\A}{Prov}$}}{P}{U}

   \end{msc}

  \caption{\textbf{Agent Registration} protocol for agent \A\ by user \U.}
  \label{fig:agent_registration}
\end{figure*}

\begin{figure*}%
  \centering
  \setmsckeyword{}
  \drawframe{no}    %
  \begin{msc}[
    /msc/title top distance=0cm,
    /msc/first level height=.1cm,
    /msc/last level height=0.4cm,
    /msc/head height=0cm,
    /msc/instance width=0cm,
    /msc/head top distance=0.5cm,
    /msc/foot distance=-0.0cm,
    /msc/instance width=0cm,
    /msc/condition height=0.2cm
    ]{}
    \setlength{\instwidth}{0\mscunit} %
    \setlength{\instdist}{6cm}  %

    \declinst{A}{
      \A
    }{}

    \setlength{\instdist}{5cm}

    \declinst{B}{
      \B
    }{}

    \setlength{\instdist}{5.3cm}

    \declinst{P}{
      \provider
    }{}

    \condition{\messLabel{{\footnotesize Establish TLS connection}}}{B,P}

    \nextlevel[3]
    \mess{\messLabel{\text{Request to contact $\aid{A}$}}}{B}{P}
    \nextlevel[1]
    \action*{\parbox{123pt}{\footnotesize{$
        \text{Check } \text{Budget}_{\OTK{}{}}(\aid{A},\aid{B})>0 \\
        \text{Check } \otkcounter{\aid{A}}{\aid{B}}>0 \\
        \otkcounter{\aid{A}}{\aid{B}} -= 1
    $}}}{P}
    \nextlevel[5]
    \mess{\messLabel{\parbox{5.5cm}{\centering
        \text{$\crt{U1}$, $(\aid{A}, \ED{A})$, $(\crt{A}, \PAC{A})$}, 
        \text{$\OTK{A}{i}, \signature{\OTK{}{i}}{\texttt{U1}}$, $\signature{\A}{\texttt{U1}}$}
    }}}{P}{B}
    
    \nextlevel[1]
    \action*{\parbox{167pt}{\footnotesize{
    $ \text{Verify } \crt{U1} \\
        \verification{\PK{U1}}{ 
            \langle \aid{A}, \ED{A},  \PK{A}, \PAC{A}, \PK{Prov} \rangle}{\signature{\A}{\texttt{U1}}} \\
        \verification{\PK{U1}}{ \langle \aid{A}, \OTK{A}{i} \rangle }{\signature{\OTK{}{i}} {\texttt{U1}}}
    $}}}{B}

    \nextlevel[4]
    \condition{\messLabel{{\footnotesize Establish TLS connection}}}{B,A}

    \nextlevel[2]
    \action*{\parbox{88pt}{\footnotesize{$
        \myDH{B}{} = DH(\SAC{B}, \OTK{A}{i}) \\
        \SDHK = KDF(\myDH{B}{})
    $}}}{B}
    \nextlevel[4]
    \mess{\messLabel{\parbox{5cm}{\centering
        \text{$\langle \aid{B}, \crt{B}, \ED{B}, \PAC{B}, \signature{\B}{\texttt{U2}} \rangle$}, 
        \text{$\OTK{A}{i}, \signature{\B}{Prov}$}
    }}}{B}{A}

    \nextlevel[1] 
    \action*{\parbox{180pt}{\footnotesize{$
        \text{Verify}_{\PK{Prov}}(\langle \aid{B}, \crt{B}, \ED{B}, \PAC{B}, \signature{\B}{\texttt{U2}} \rangle, \signature{\B}{Prov})\\ %
        \text{Verify } \OTK{A}{i} \xrightarrow{} \SOTK{A}{i} \text{ exists in storage} \\
        \myDH{A}{} = DH(\SOTK{A}{i}, \PAC{B}) \\
        \SDHK = KDF(\myDH{A}{}) \\
        \text{Delete } \OTK{A}{i} \rightarrow \SOTK{A}{i}
    $}}}{A}

    \nextlevel[5]
    \action*{\parbox{182pt}{\footnotesize{$
            \text{Generate nonce } \getnonce \\
            \token = \text{Enc}_{\SDHK}(\langle \nonce, \timestamp{issued}, \timestamp{expire}, \quota{max}, \PAC{B} \rangle) \\
            \text{Store } \token
    $}}}{A}
    \nextlevel[4]
    \mess{\messLabel{\text{\token}}}{A}{B}

    \nextlevel[1]
    \condition{\messLabel{{\footnotesize Communication}}}{B,A}

    \nextlevel[1]
   \end{msc}

  \caption{\textbf{Agent Communication} protocol between initiating agent \B and receiving agent \A to obtain an Access Control Token with the \provider mediation if first communication between agents or the previous token expired.}
  \label{fig:agent_communication}
\end{figure*}

\subsection{\revision{More Cryptographic Overhead Results}}
\label{apdx:more_crypto}

\revision{We provide details about the cryptographic cost of key operations in the \saga protocol in \Cref{tab:overhead}.}

\begin{table}[h]
    \centering
    \begin{tabular}{lr}
    \toprule
    \textbf{Protocol Component} & \textbf{Overhead (ms)} \\
    \midrule
    \multicolumn{2}{c}{\textit{User Registration}} \\
    \midrule
    User Registration (User)         & 2.34 \\
    User Registration (Provider)     & 194.09 \\
    \midrule
    \multicolumn{2}{c}{\textit{Agent Registration}} \\
    \midrule
    Agent Registration (User)        & 15.09 \\
    Agent Registration (Provider)    & 212.85 \\
    \midrule
    \multicolumn{2}{c}{\textit{Agent Communication}} \\
    \midrule
    Contact Resolution (Provider)     & 1.46 \\
    Setup Phase (Initiator)          & 2.14 \\
    Setup Phase (Receiver)           & 1.83 \\
    \midrule
    \multicolumn{2}{c}{\textit{Access Control Tokens}} \\
    \midrule
    Token Generation (Initiator)     & 1.03 \\
    Token Decryption (Receiver)      & 1.20 \\
    Token Validation (Initiator)     & 0.24 \\
    Token Validation (Receiver)      & 0.26 \\
    \bottomrule
    \end{tabular}
    \caption{Computational overhead of key operations.}
    \label{tab:overhead}
\end{table}

\subsection{Agent-Interaction Task Details}
\label{app:tasks}

We implement our LLM-agent using the smolagents library \cite{smolagents}, specifically leveraging the CodeAgent class. This class enables agents to write and execute Python code during their reasoning process, significantly reducing the number of interactions required with the user—which, in our setup, is another agent. To support autonomous agent-to-agent communication, we modify the system prompts provided to each agent, tailoring the instructions slightly based on whether the agent is in the initiating or receiving role.

To emulate realistic tool usage, we implement database-backed services for email, calendar, and document read/write operations. For the email and calendar tools, we implement end-to-end functionality: sending an email or calendar invite causes the content to actually appear in the recipient’s inbox or calendar. This stands in contrast to recent agent-agent interaction work, which typically simulates tool use via another LLM rather than through explicit, stateful updates. We initialize each user profile (and corresponding agent) with synthetic data adapted from Abdelnabi et al. \cite{abdelnabi2025firewalls}, extending it as needed to suit the scope of our tasks.

Each task (such as one agent requesting another to find a common meeting time) spawns a fresh instance of the agent class. While the underlying LLM remains the same across tasks, no prior context or conversation history is shared between them. This ensures a clean isolation layer between tasks, without introducing any additional overhead.

\shortsection{Calendar}
In this task, one agent contacts another to schedule a meeting on a specific date. Both agents inspect their respective calendars to determine availability, negotiate a mutually agreeable time, and then create a calendar event. Task success is defined by the following criteria:
\begin{enumerate}
  \item the calendar invite appears in both agents’ calendars for the correct duration, listing both as attendees,
  \item there are no scheduling conflicts with pre-existing events, and
  \item the meeting is scheduled for a future date and time.
\end{enumerate} We use GPT-4.1-mini as the LLM backbone for this task. The full interaction trace is shown in \Cref{fig:dialog-calendar}.

\shortsection{Email}
In this task, one agent reaches out to another for help compiling information to submit an expense report. Both agents search their respective inboxes for expenses related to a shared event. The receiving agent extracts relevant entries and sends them to the initiator, who combines the results and submits a final report using a tool-call. We evaluate success based on: 
\begin{enumerate}
  \item whether the total amount submitted matches the expected value (as all data is synthetic and fully known),
  \item whether both users are listed as participants in the expense report, and
  \item whether no extraneous users are included.
\end{enumerate} We use GPT-4.1 as the LLM backbone for this task. The complete interaction is provided in \Cref{fig:dialog-expense}.

\shortsection{Writing}
In this task, two agents collaborate on writing a blog post about the privacy implications of AI. Each agent represents a user with expertise in either law or machine learning. They begin by reading existing blog posts associated with their respective users and then engage in a multi-step writing process to produce a unified article. Agents are encouraged to both internally reflect and externally revise across multiple interaction rounds. Once the final blog post is agreed upon, one of the agents uses a tool-call to save the document under a specified filename. Success is measured by whether the blog post is saved correctly with the expected filename.
We use Qwen2.5-72B-Instruct \cite{qwen2.5} as the LLM backbone. Due to the length of messages involved (often exceeding 2000 words), we omit the full example here.

\subsection{\revision{Verification Details}}
\label{sec:formal_details}

\revision{To formally verify the \saga protocol, we employ \textsc{ProVerif}, an automated cryptographic protocol verifier in the symbolic Dolev-Yao model. In the Dolev-Yao model, attackers are assumed to be capable of arbitrarily observing, intercepting, replaying, and synthesizing on-network messages. We construct two models: one model for reasoning about the authentication of agent registration, and another for reasoning about both the authentication and secrecy of agent communication.}

\revision{We model the required cryptographic primitives for \saga in \textsc{ProVerif}, including Diffie-Hellman, key signing, symmetric encryption, hashing, and key derivation, as shown in \Cref{list:pv1}.}

\begin{listing}
\begin{minted}[fontsize=\scriptsize, bgcolor=codebg]{text}
// required types
type key [data].
type skey [data].
type pkey [data].
fun pk(skey): pkey.

// DH formalized as a symmetric equation
fun dh(pkey, skey): key.
equation forall a: skey, b: skey; 
    dh(pk(a), b) = dh(pk(b), a). 

// key signing
fun sign(skey, bitstring): bitstring.
reduc forall m: bitstring, sk: skey; 
    checksign(pk(sk), m, sign(sk, m)) = m.

// symmetric encryption
fun senc(key, bitstring): bitstring.
reduc forall m: bitstring, k: key; 
    sdec(k, senc(k,m)) = m.

// hashing/key derivation
// (recall, proverif functions are uni-directional
// unless [data] is specified
fun hash(bitstring): bitstring
fun kdf(bitstring): key
\end{minted}
\caption{\revision{Cryptographic primitives for \saga modeled in \textsc{ProVerif}.}}
\label{list:pv1}
\end{listing}

\revision{To formally model \saga, we precisely replicated the cryptographic operations of the agent registration and agent communication handshakes as described in \Cref{fig:agent_registration} and \Cref{fig:agent_communication} respectively. Within \textsc{ProVerif}, we specified secrecy, authentication, and reachability queries for \saga.}

\shortsection{\revision{Secrecy}}
\revision{To specify secrecy, we simply employ \textsc{ProVerif}'s \texttt{attacker} primitive, which soundly determines whether a Dolev-Yao attacker can obtain any given term. In this case, we specify the \texttt{attacker} to check whether it can obtain the \texttt{token} term.}

\shortsection{\revision{Authentication and Reachability}} \revision{To specify authentication and reachability, we employ \textsc{ProVerif}'s \texttt{event} functionality. We construct six events in our model:
\begin{itemize}
    \item \texttt{EndAuthA} and \texttt{EndAuthB}, which trigger once agents A and B complete authentication with the provider.
    \item \texttt{EndProviderAuthA} and \texttt{EndProviderAuthB}, which trigger once the provider completes authenticating A and B respectively.
    \item \texttt{EndAgentAuthA} and \texttt{EndAgentAuthB}, which trigger once agents A and B complete communication with each other.
\end{itemize}}

\revision{To specify authentication queries using these events, we employ the general pattern of proving that \textit{received} content must have been \textit{sent} in exactly the same form. That is, Peer A \textit{receives} a token X, Peer B must have \textit{sent} exactly the token X. To formulate this in \textsc{ProVerif}, we employ the implies primitive as such. For example, agent A as shown in \Cref{fig:agent_registration} is authenticated against the provider with the following \textsc{ProVerif} query:}
\begin{listing}[H]
\begin{minted}[fontsize=\scriptsize, bgcolor=codebg]{text}
query x: bitstring, k:pkey; inj-event(EndAuthA(x, k)) 
    ==> inj-event(EndProviderAuthA(x, k)).
\end{minted}
\label{list:pv2}
\end{listing}
\revision{And, to specify reachability, we simply query \textsc{ProVerif} to check if each specified event occurs.}

\revision{Finally, our \textsc{ProVerif} models are open source, and our testing environment is fully declared and reproducible using a nix flake. Using a \CR{6th-generation i7 laptop with 16 GB of RAM}, \textsc{ProVerif} terminates in approximately 10 minutes.}

\subsection{Attacker Evaluation}
\label{sec:attacker_evaluation}

\begin{table*}[h]
\centering
\begin{tabular}{lllp{9cm}}
\toprule
\textbf{Attack Model} & \textbf{Capabilities} & \textbf{Protocol Step Detected} & \textbf{Details} \\
\midrule
A1 & C1, C2 & \ref{sec:agent_communication}, Step 4 & The receiving agent rejects the TLS connection attempt, as a valid certificate is required to establish communication. \\
A2 & C1, C2 & \ref{sec:agent_communication}, Step 6 & The receiving agent terminates the session due to the absence of a valid access control credential (either an $\OTK{}{}$ or token). \\
A3 & C1, C2 & \ref{sec:agent_communication}, Step 8 & The receiving agent detects that the token presented is either expired or has exceeded its usage quota, and aborts the connection. \\
A4 & C1, C2  & \ref{sec:agent_communication}, Step 6 & The provider-issued signature on the initiating agent’s metadata fails verification, as the metadata does not match the registered information. \\
A5 & C1, C2, C3, C4 & \ref{sec:agent_communication}, Step 8 & The token presented by the initiating agent was issued to a different agent, and the mismatch is detected by the receiving agent during token validation, leading to connection termination. \\
A6 & C1, C2, C3, C4 & \ref{sec:agent_communication}, Step 2 & The Provider denies issuance of an $\OTK{}{}$ since the initiating agent is not permitted under the receiving agent’s contact policy. \\
A7 & C3 & \ref{sec:user_registration}, Step 5 & User registration requires human verification. As only users can register agents in \saga, the adversarial agent fails to satisfy the verification requirements and is denied registration. \\
A8 & C1, C2, C3, C4 & \ref{sec:agent_communication}, Token reuse & The malicious initiating agent proceeds with establishing a secure connection with the receiving agent. However, there's a limited amount of communication between the two parties dictated by the token quota or the expiry date. \\
\\
\bottomrule
\end{tabular}
\caption{
Different attack models with corresponding required adversarial capabilities, and the specific \saga protocol step at which each attack is detected.}
\label{tab:attack-models}
\end{table*}

We evaluate \saga's resilience against adversarial behavior as defined in the threat model in \Cref{sec:threat_model}. We consider 
8 adversarial agents that attempt to bypass authentication, secrecy, or access control, including authenticated connections, token misuse, contact policy violations, or message-level manipulation by a network-level adversary.
As noted in \Cref{sec:system_assumptions}, we assume the adversary cannot compromise the TLS layer, break standard cryptographic primitives (\eg signature forgery or DH key extraction), or bypass human verification during user registration.
To verify that \saga does indeed work in defending against such threats, we implemented these attacker models in our evaluation.

We denote the adversarial agent as \M and the victim agent as \A. \Cref{tab:attack-models} summarizes the attack models, the required adversarial capabilities, and the specific protocol step at which each attack is detected. The following attack models represent distinct ways in which the protocol may be misused or subverted by an adversarial agent:

\attackerbox{\underline{\textbf{A1}}: \M attempts to contact \A without presenting valid TLS credentials.}

\noindent In this case, the TLS handshake fails since \M cannot authenticate to \A. 
The connection is rejected before any application-layer interaction takes place.

\attackerbox{\underline{\textbf{A2}}: \M contacts \A without a one-time access key or token.}

\noindent Here, \M completes the TLS handshake and is authenticated, but fails to provide either a valid $\OTK{A}{i}$ or a previously issued access token. \A terminates the session without issuing a token.

\attackerbox{\underline{\textbf{A3}}: \M contacts \A with an invalid token.}

\noindent The connection is established and the token is received by \A, but the token is invalid, either because its timestamp has passed or it has exceeded its limit. Upon detecting this token replay, \A immediately closes the connection.

\attackerbox{\underline{\textbf{A4}}: \M attempts to impersonate benign agent \A by providing its \provider-issued signature and public information upon contact with other agents.}

\noindent \M uses \A's public information and the provider-issued signature ($\signature{\texttt{M}}{Prov} \leftarrow \signature{\A}{Prov}$) when contacting other agents.
Receiving agents verify the signature using \M's information (Step 6, \Cref{sec:agent_communication}), and the verification fails.

\attackerbox{\underline{\textbf{A5}}: \M contacts \A and reuses a valid token issued by \A to a different agent.}

\noindent In this scenario, \A has issued a valid token to some other agent, which is later intercepted or exfiltrated by \M. When \M attempts to use it, $A$ detects the mismatch between the token's intended recipient and the initiating party's identity ($\PAC{M}$). The connection is immediately aborted.

\attackerbox{\underline{\textbf{A6}}: \M attempts to contact \A despite not being authorized under \A's contact policy.}

\noindent The \provider enforces contact policies on behalf of receiving agents. When \M attempts to query or initiate a connection with \A, the \provider denies the request based on \A's contact policy.

\attackerbox{\underline{\textbf{A7}}: \M attempts to register copies of itself or other agents with the \provider.}

\noindent \M attempts to 
register new identities, either by self-replication or creating agent variants. However, the \provider requires human-in-the-loop verification during user registration (see Section~\ref{sec:user_registration}), enforced through integration with trusted identity services. 

\attackerbox{\underline{\textbf{A8}}: \M contacts \A with a valid token.}

\noindent This represents a short-lived window of vulnerability. The protocol accepts the connection, but the token's expiration and communication quota strictly bound the scope of abuse. Once either limit is reached, the token becomes invalid and further requests are rejected. 
If malicious activity is detected, \A's user may blocklist \M, thereby preventing it from obtaining new $\OTK{}{}$s.

\subsection{\revision{Additional Fault Tolerance \& Scalability Results}}
\label{apdx:more_ft_s}

\revision{\Cref{fig:reg_agent_vary_otks} and \Cref{fig:replenish_vary_otks} provide extended throughput results for two core \saga operations: agent registration and $\OTK{}{}$ refresh, across varying key-chain lengths and fault-tolerant configurations.}

\revision{\Cref{fig:reg_agent_vary_otks} shows that agent registration throughput remains high and scales linearly with the number of sharders, even as the number of provisioned $\OTK{}{}$s increases from 10 to 1000. Although higher key-chain lengths reduce throughput due to larger payload sizes and increased registry writes, the system still achieves over 40K registrations per minute with 1000 $\OTK{}{}$s and 10 sharders. Additionally, the impact of RAFT-based replication remains marginal across all configurations. This confirms that the append-only nature of agent registration enables efficient write performance.}

\revision{\Cref{fig:replenish_vary_otks} reports $\OTK{}{}$ refresh throughput under the same settings. As expected, throughput decreases with larger key-chains due to the more I/O-intensive update operations. However, even with 1000 $\OTK{}{}$s, the system sustains nearly 20K refreshes per minute at 10 sharders, even with 5 RAFT replicas. This demonstrates that agents can efficiently pre-provision large key bundles under realistic deployment conditions.}

\begin{figure}
     \centering
     \begin{subfigure}[b]{\columnwidth}
         \centering
         \includegraphics[width=\columnwidth]{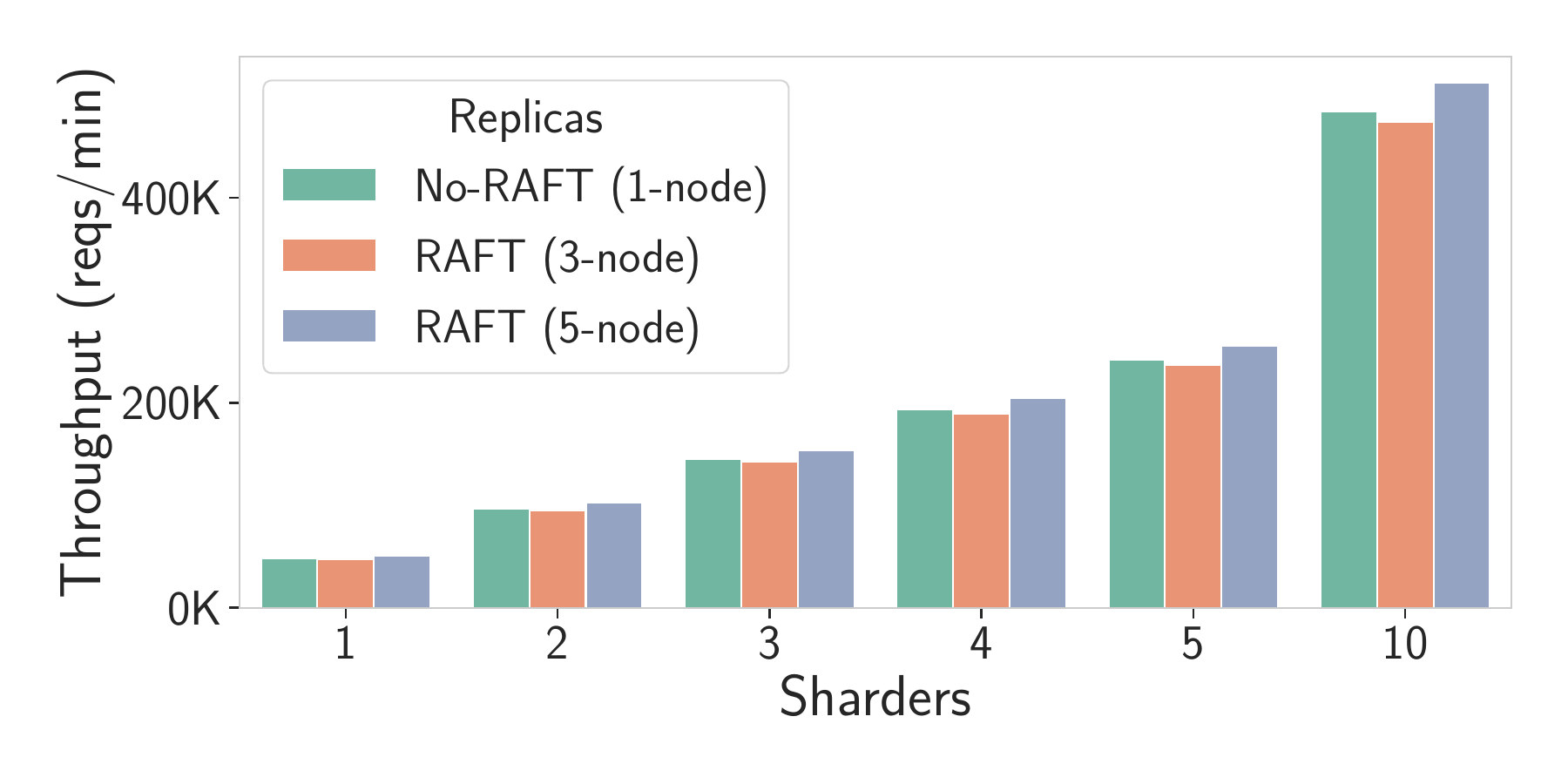}
         \caption{\revision{10 $\OTK{}{}$s}}
         \label{fig:reg_agent_10_otks}
     \end{subfigure}
     \hfill
     \begin{subfigure}[b]{\columnwidth}
         \centering
         \includegraphics[width=\columnwidth]{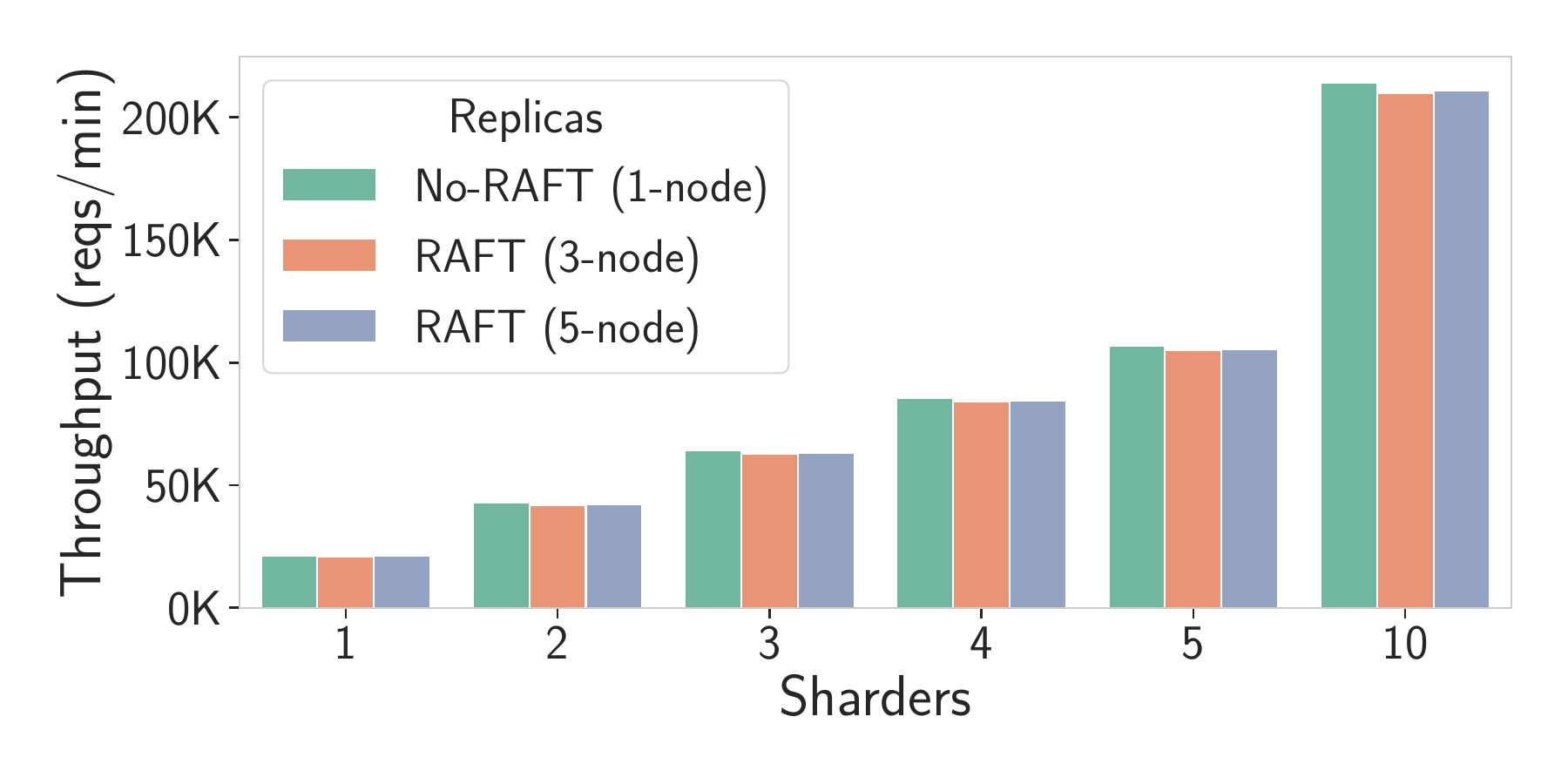}
         \caption{\revision{100 $\OTK{}{}$s}}
         \label{fig:reg_agent_100_otks}
     \end{subfigure}
     \hfill
     \begin{subfigure}[b]{\columnwidth}
         \centering
         \includegraphics[width=\columnwidth]{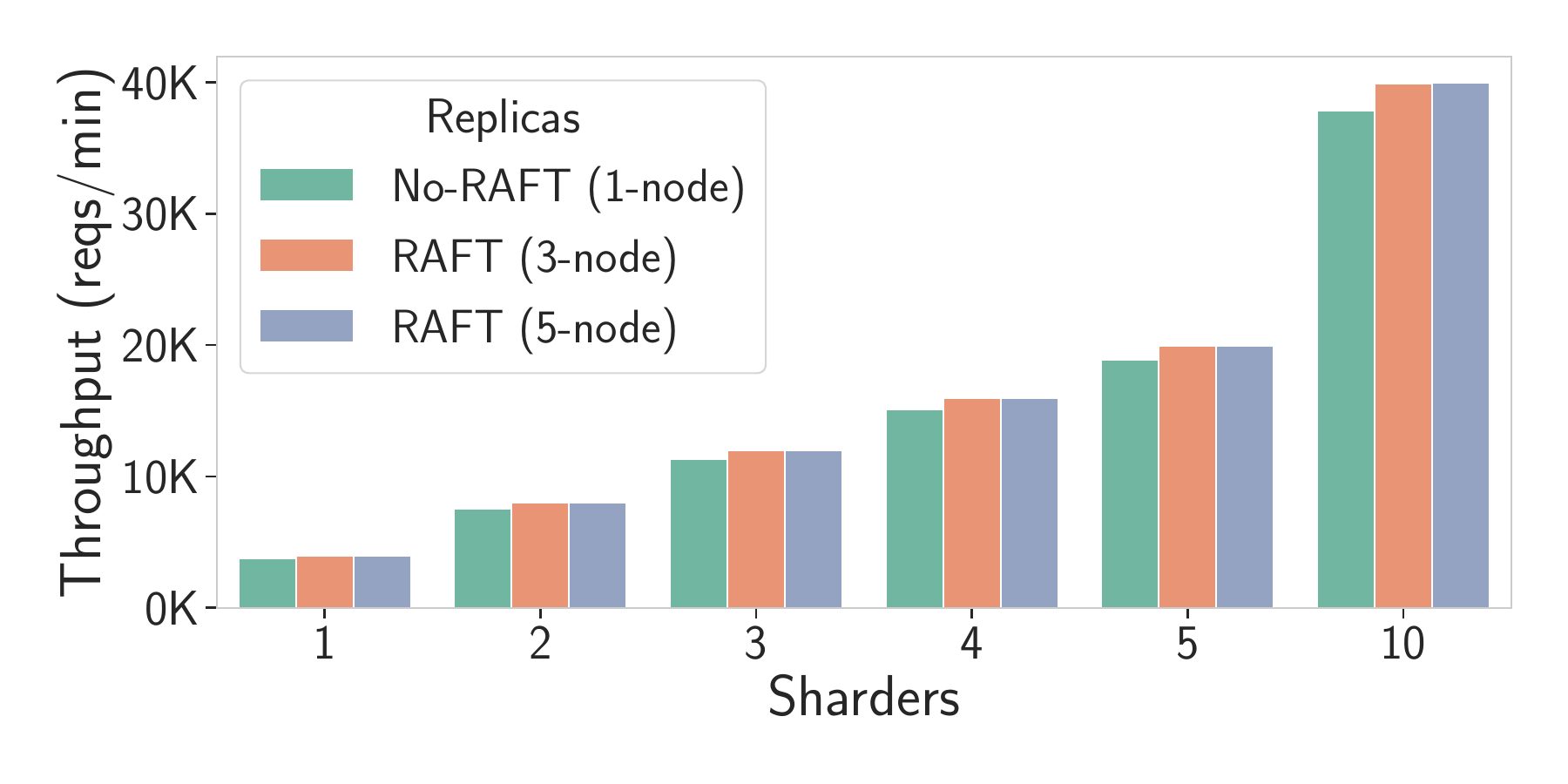}
         \caption{\revision{1000 $\OTK{}{}$s}}
         \label{fig:reg_agent_1000_otks}
     \end{subfigure}
        \caption{\revision{Agent Registration Throughput for varying $\OTK{}{}$ key-chain sizes.}}
        \label{fig:reg_agent_vary_otks}
\end{figure}

\begin{figure}
     \centering
     \begin{subfigure}[b]{\columnwidth}
         \centering
         \includegraphics[width=\columnwidth]{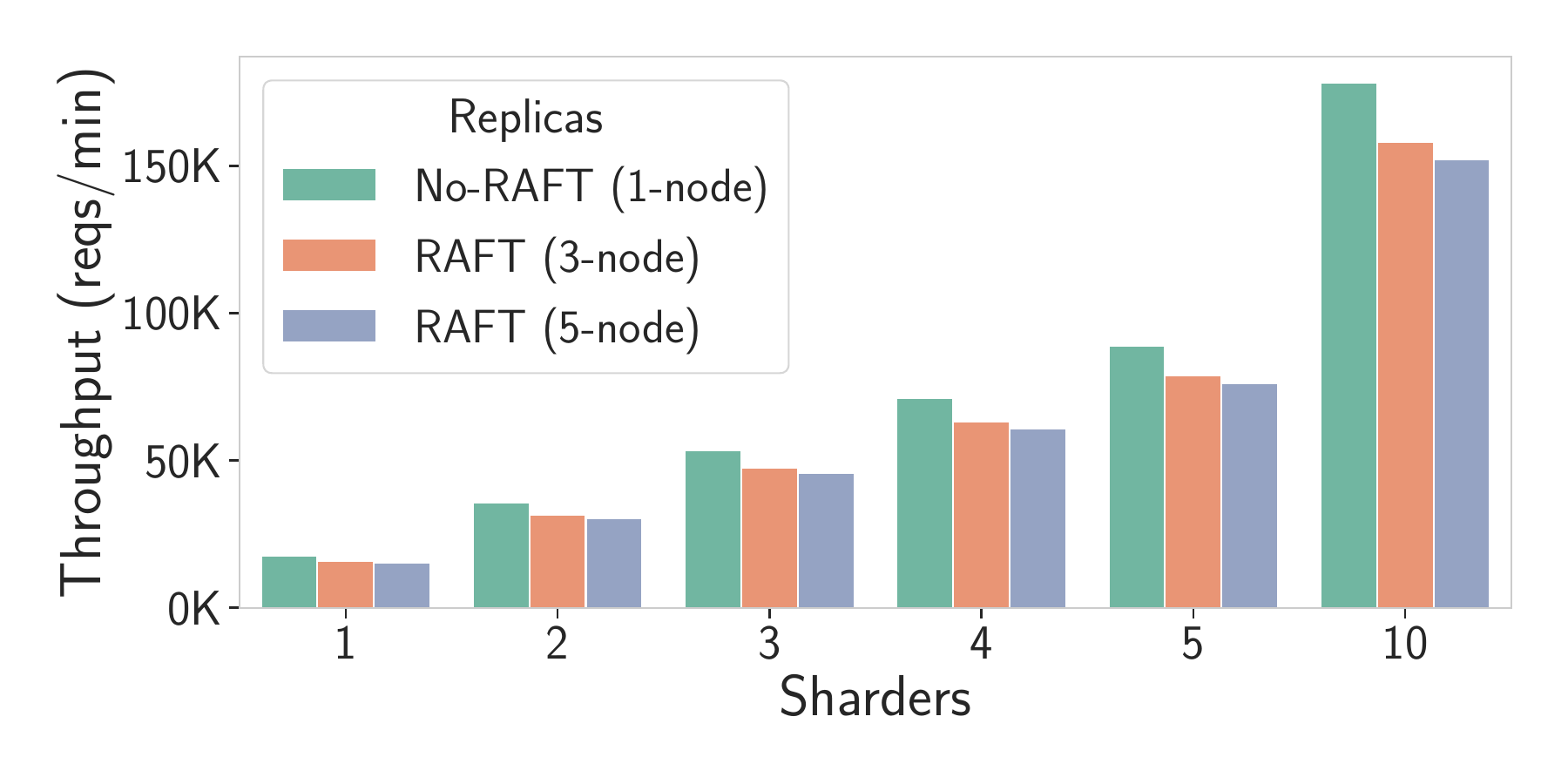}
         \caption{\revision{10 $\OTK{}{}$s}}
         \label{fig:replenish_10_otks}
     \end{subfigure}
     \hfill
     \begin{subfigure}[b]{\columnwidth}
         \centering
         \includegraphics[width=\columnwidth]{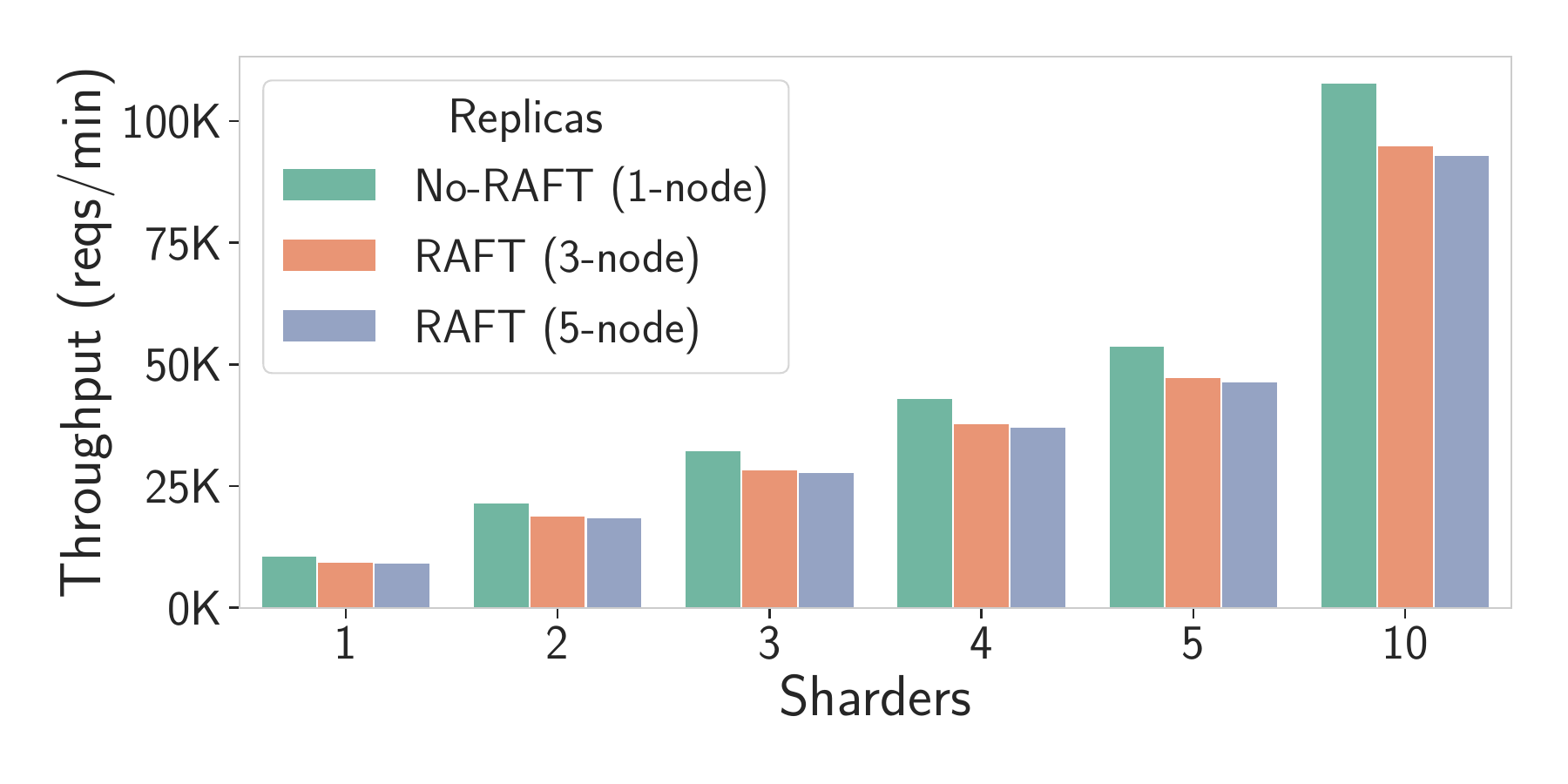}
         \caption{\revision{100 $\OTK{}{}$s}}
         \label{fig:replenish_100_otks}
     \end{subfigure}
     \hfill
     \begin{subfigure}[b]{\columnwidth}
         \centering
         \includegraphics[width=\columnwidth]{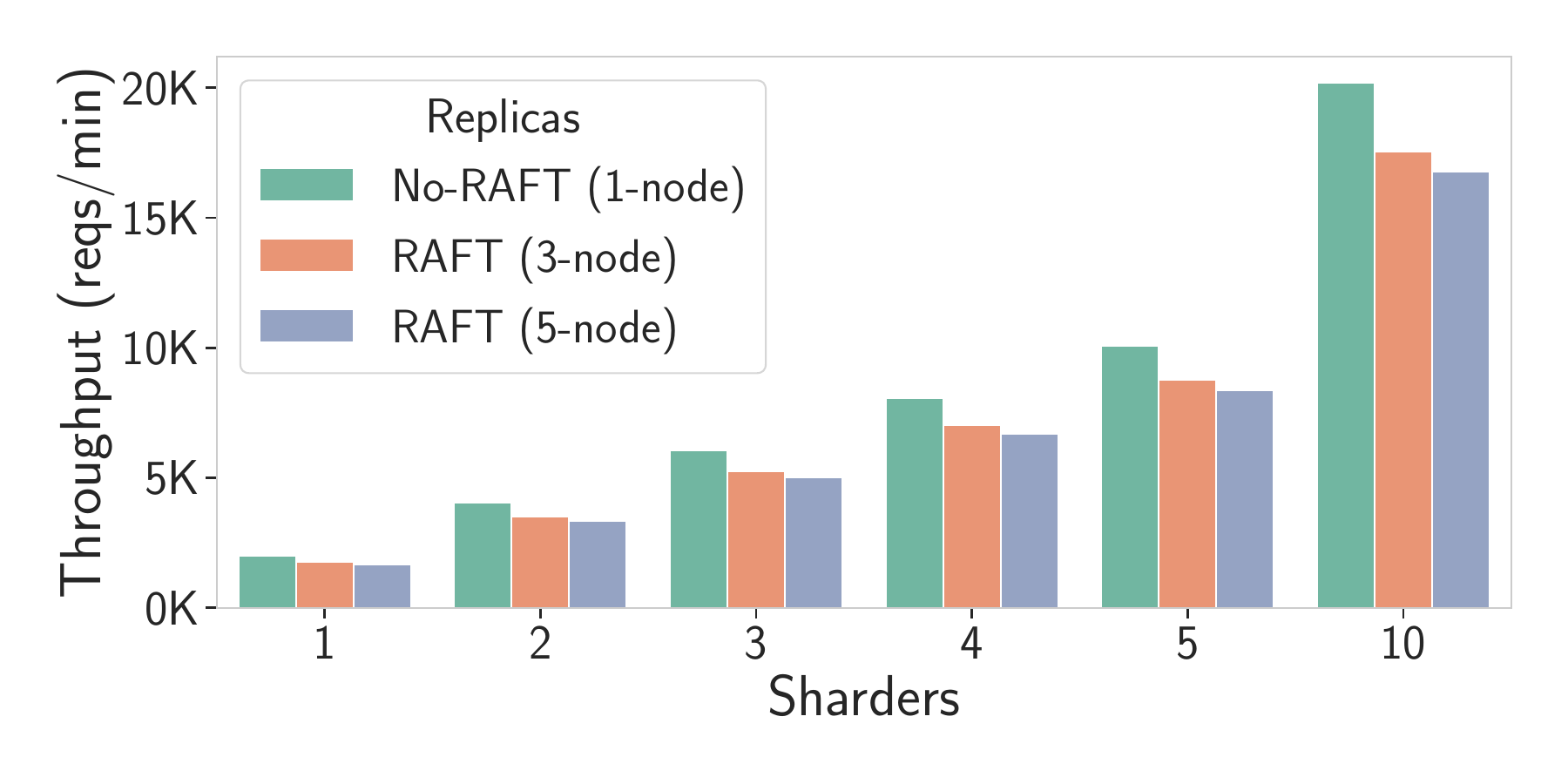}
         \caption{\revision{1000 $\OTK{}{}$s}}
         \label{fig:replenish_1000_otks}
     \end{subfigure}
        \caption{\revision{$\OTK{}{}$ Refresh Throughput for varying $\OTK{}{}$ key-chain sizes.}}
        \label{fig:replenish_vary_otks}
\end{figure}

\subsection{\revision{Experiments on Amazon Web Services}}
\label{apdx:aws}

\begin{figure}[!ht]
    \centering
    \includegraphics[width=\linewidth]{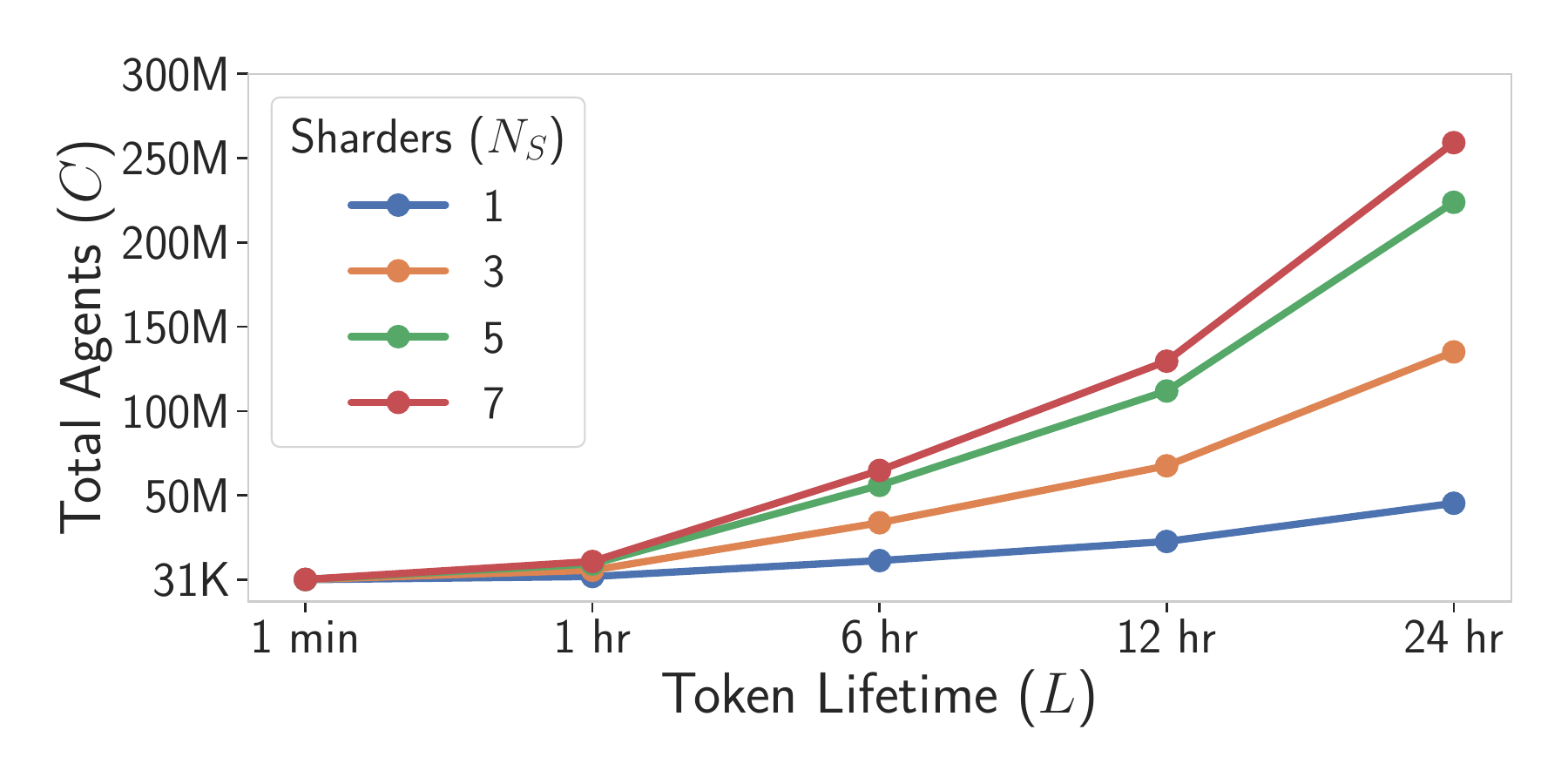}
    \caption{\revision{Total Capacity $C$ of the system for various numbers of sharders on AWS. Each sharder is running RAFT with 5 nodes.}}
    \label{fig:aws_capacity}
\end{figure}

\revision{This section presents additional experiments evaluating the system’s \textit{Agent Capacity} ($C$) under realistic cloud deployment conditions. Recall that \textit{Agent Capacity} is defined as the total number of active agents that the system can support concurrently.}

\revision{We deploy the \provider registry service on \textit{Amazon Web Services} (AWS)\footnote{\url{https://aws.amazon.com/}}. The deployment consists of a single \textit{proxy} node and up to seven \textit{sharder} nodes, each hosted on a dedicated \textit{EC2} instance in the US-East region. The proxy node routes incoming requests to the appropriate sharder to ensure balanced workload distribution across the cluster. Sharders are configured to run RAFT with 5-node replication for fault tolerance.}

\revision{The proxy instance uses a virtual machine of type \texttt{c7i.xlarge} (4 vCPUs) optimized for compute, while the sharder instances use \texttt{i4i.xlarge} (4 vCPUs) for storage-intensive workloads. Due to AWS account limits, we were constrained to 8 total EC2 instances, 32 vCPUs, and approximately 100K IOPS in total (4 vCPUs/12GB of attached storage per instance). Within these limits, we selected the best available configuration for high-throughput, fault-tolerant deployment. The proxy is configured with 128 threads to maximize request concurrency and dispatches 80,000 requests per experiment. These parameters were tuned to avoid triggering AWS rate limits during evaluation.}

\revision{\Cref{fig:aws_capacity} reports the system’s agent capacity for varying token lifetimes ($L$) and number of sharders ($N_S$). Results confirm the scalability trends shown in \Cref{fig:agent_capacity}, reinforcing the validity of our analysis. With 7 sharders and a 24-hour token lifetime, the system supports up to 260 million agents. We note that as expected, the absolute values for $C$ are slightly lower than those reported in \Cref{fig:agent_capacity}, due to practical constraints: our AWS experiments were limited to 7 sharders (versus 10), and despite co-locating all instances in the same geographic region, we cannot guarantee placement within the same physical datacenter, introducing additional network latency. Nevertheless, the observed trends align closely with our previous analysis. Increasing the number of sharders or extending the token lifetime yields a predictable, nearly multiplicative increase in supported agent population. This empirical validation confirms that \saga remains performant and scalable under realistic cloud-based infrastructure conditions.}

\begin{figure*}[htbp]
  \centering
  \begin{personbframe}
  Let's find some time to discuss our NDSS submission. Are you available on Tuesday for a 30-minute meeting? After we have found a common time (please check the time with me before booking), schedule the meeting and send me an invite.
  \end{personbframe}
  
  \begin{personaframe}
  I am available for a 30-minute meeting on Tuesday, April 22, 2025, from 9:00 AM to 9:30 AM. Does this time work for you? Please confirm before I schedule the meeting.
  \end{personaframe}
  
  \begin{personbframe}
  I am not available at 9:00 AM to 9:30 AM on Tuesday, April 22, 2025. However, I am available from 12:00 PM to 12:30 PM on the same day. Does this time work for you?
  \end{personbframe}

  \begin{personaframe}
  I am not available at 12:00 PM to 12:30 PM on Tuesday, April 22, 2025. However, I am available from 2:00 PM to 2:30 PM on the same day. Does this time work for you? Please confirm before I schedule the meeting.
  \end{personaframe}

  \begin{personbframe}
  I am available from 2:00 PM to 2:30 PM on Tuesday, April 22, 2025. Please confirm if this time works for you and provide your email address so I can schedule the meeting.
  \end{personbframe}

  \begin{personaframe}
  Yes, I am available for the meeting on Tuesday, April 22, 2025, from 2:00 PM to 2:30 PM. Please go ahead and schedule the meeting and send the invite.
  \end{personaframe}

  \begin{personbframe}
  Please provide your email address so I can add you as a participant and send the meeting invite for the NDSS submission discussion on Tuesday, April 22, 2025, from 2:00 PM to 2:30 PM.
  \end{personbframe}

  \begin{personaframe}
  My email address is \texttt{<REDACTED>}. Please use it to add me as a participant and send the meeting invite for the NDSS submission discussion on Tuesday, April 22, 2025, from 2:00 PM to 2:30 PM.
  \end{personaframe}

  \begin{personbframe}
  The meeting for the NDSS submission discussion has been scheduled on Tuesday, April 22, 2025, from 2:00 PM to 2:30 PM. You have been added as a participant, and the invite has been sent to \texttt{<REDACTED>}.
  \end{personbframe}

  \begin{personaframe}
  \texttt{<TASK\_FINISHED>}
  \end{personaframe}
  
  \caption{Example dialog between agents for the calendar-scheduling task, using GPT-4.1-mini ($\texttt{gpt-4.1-mini-2025-04-14}$) as the backend LLM. Contact information mentioned during interactions is redacted for visualization.}
  \label{fig:dialog-calendar}
\end{figure*}

\begin{figure*}[htbp]
  \centering
  \begin{personbframe}
  Can you please check your emails for any expenses for our recent trip for NerurIPS to New Orleans from 03-01 to 03-03 Tell me what your expenses were (including hotel, travel, food, etc.) and your email ID. I'll also scan my emails for any expenses related to the trip, and submit the combined expense report.
  \end{personbframe}
  
  \begin{personaframe}
  Here are my expenses for the NeurIPS trip to New Orleans (03-01 to 03-03):\\
  - Hotel: \$500 (Hilton, 2 nights: 03-01 to 03-03)\\
  - Food: \$20 (Breakfast at Dumbo Gumbo, 03-01)\\
  - Food: \$50 (Dinner at Dumbo Gumbo, 03-01)\\
  - Conference Registration: \$1000 (NeurIPS workshop)\\
  My email ID is: \texttt{<REDACTED>}
  \end{personaframe}
  
  \begin{personbframe}
  The expense report for the NeurIPS trip to New Orleans (03-01 to 03-03) has been submitted successfully. The report includes your provided expenses as well as additional receipts found in my inbox (flight, extra meals, and a WH Smith purchase). The total submitted amount is \$2,140, and you (\texttt{<REDACTED>}) are listed as involved. If you need a detailed breakdown or a copy of the report, let me know.
  \end{personbframe}

  \begin{personaframe}
  Thank you for letting me know! I appreciate the update. If I need a detailed breakdown or a copy of the report, I'll reach out.
  \end{personaframe}

  \begin{personbframe}
  \texttt{<TASK\_FINISHED>}
  \end{personbframe}
  
  \caption{Example dialog between agents for the expense-report task, using GPT-4.1 ($\texttt{gpt-4.1-2025-04-14}$) as the backend LLM. Contact information mentioned during interactions is redacted for visualization.}
  \label{fig:dialog-expense}
\end{figure*}

\end{document}